\documentclass[prd,
,superscriptaddress,showpacs,nofootinbib,%
tightenlines ]{revtex4}
\usepackage{epsfig}
\usepackage{amsfonts}
\usepackage{amssymb}

\newcommand{\ben}{\begin{displaymath}}
\newcommand{\een}{\end{displaymath}}
\newcommand{\be}{\begin{equation}}
\newcommand{\ee}{\end{equation}}
\newcommand{\bea}{\begin{eqnarray}}
\newcommand{\eea}{\end{eqnarray}}
\begin{document}
\title{Vacuum energy in the effective field theory of general relativity}

\author{J.~Gegelia}
\affiliation{Ruhr University Bochum, Faculty of Physics and Astronomy,
Institute for Theoretical Physics II, D-44870 Bochum, Germany}
\affiliation{Tbilisi State  University,  0186 Tbilisi,
 Georgia}
\author{Ulf-G.~Mei{\ss}ner}
 \affiliation{Helmholtz Institut f\"ur Strahlen- und Kernphysik and Bethe
   Center for Theoretical Physics, Universit\"at Bonn, D-53115 Bonn, Germany}
 \affiliation{Institute for Advanced Simulation, Institut f\"ur Kernphysik
   and J\"ulich Center for Hadron Physics, Forschungszentrum J\"ulich, D-52425 J\"ulich,
Germany}
\affiliation{Tbilisi State  University,  0186 Tbilisi, Georgia}

\begin{abstract}

In the framework of an effective field theory of general relativity a model of scalar and vector bosons interacting with the
metric field is considered.  It is shown in the framework of a two-loop order calculation 
that for the cosmological constant term which is fixed by the condition of vanishing vacuum energy the graviton remains massless and
there exists a self-consistent effective field theory of general relativity coupled to matter fields defined on a
flat Minkowski background. This result is obtained under the assumption that the
energy-momentum tensor of the gravitational field  is given by the pseudotensor of Landau-Lifshitz's classic textbook.
Implications for the cosmological constant problem are also briefly discussed.

\end{abstract}


\maketitle

\section{Introduction}	
	
It is widely accepted that whatever the underlying fundamental theory of all interactions might be at low energies,
the physics can be adequately described by an effective field theory (EFT) \cite{Weinberg:1995mt}. 
Gravitation can also included in the formalism of EFT  by considering the most general effective Lagrangian of metric
fields interacting with matter fields  \cite{Donoghue:1994dn,Donoghue:2015hwa} which is invariant under all underlying
symmetries including the gauge symmetry of massless spin-two particles \cite{Veltman:1975vx}. 
This quantum field theoretical treatment of general relativity with the metric field presented as the Minkowski
background plus the graviton field and the cosmological constant usually set equal to zero is considered as a
well-defined approach in the modern sense,  see, e.g., Ref.~\cite{Donoghue:2017pgk}.
It is well-known that for a non-vanishing cosmological constant term $\Lambda$ the graviton propagator has a pole corresponding
to  a massive ghost mode \cite{Veltman:1975vx}. 
Setting $\Lambda$ equal to zero as is usually done in the EFT of gravitation \cite{Donoghue:1994dn} does not solve
the problem, as the radiative corrections re-generate the problem with the massive ghost \cite{Burns:2014bva}. 
This is because the cosmological constant term is not suppressed by any symmetry of the effective 
theory and therefore there is  no protection against generating such a contribution to the effective action by radiative
corrections.  However, as it has been shown in Ref.~\cite{Burns:2014bva}, one can represent the cosmological constant as a
power series in $\hbar$ and choose the  coefficients of this series such that the graviton becomes a massless spin-two particle
up to all orders in the loop expansion.  Thus, within a perturbative EFT in flat Minkowski background, the cosmological constant,
which is one of the parameters of the effective Lagrangian, is uniquely fixed. 
This does not solve the cosmological constant problem \cite{Weinberg:1988cp}
(for a recent review of the cosmological constant problem see, e.g., Ref.~\cite{Martin:2012bt}) but rather implies that
taking into account  a cosmological constant term other than obtained in 
Ref.~\cite{Burns:2014bva} necessarily requires considering an EFT in a curved background field. In this case by
imposing the equations of motion with respect to the background 
graviton field the mass term of the graviton is removed  at tree level \cite{Gabadadze:2003jq}, however,
a systematic study of the issue at higher orders in loop expansion requires an 
EFT on a curved background metric which, to the best of our knowledge, is not available yet. 

Experimental evidence of the accelerating expansion of the universe (see, e.g., Ref.~\cite{Rubin:2016iqe} and
references therein) leaves us with a very challenging problem, namely the huge discrepancy between the measured small value of
the cosmological constant and its theoretical estimation \cite{Weinberg:1988cp}. 
An important question related to this problem is if there exists any condition that uniquely fixes the value of the cosmological constant.  
It seems natural to expect  that the energy of the physical vacuum state of the theory describing the universe is exactly zero. This is the main assumption of this work and of importance for the
later discussions.
In the framework of the low-energy EFT of general relativity coupled to the fields of the Standard Model imposing such a
condition uniquely fixes the cosmological constant term as a function of other parameters of the effective Lagrangian. 
In the current work we calculate the vacuum expectation value of the full four-momentum of the gravitational and matter fields
at two-loop order in a simplified version of the Abelian model with spontaneous symmetry breaking considered in
Ref.~\cite{Burns:2014bva}. 
We obtain that as a result of a non-trivial cancellation between different diagrams the vacuum energy exactly
vanishes for the value of the cosmological constant obtained in Ref.~\cite{Burns:2014bva}, i.e.,  for the value 
which guarantees the vanishing of the graviton mass  and the  vacuum expectation value of the graviton field at two-loop order.
That is, provided that our result holds to all orders, the uniquely fixed value of the cosmological constant term,
leading to a self-consistent perturbative EFT on Minkowki 
background is obtained as a consequence of imposing the condition of vanishing vacuum energy. 
Notice here  that being aware of the lack of a commonly accepted expression of the energy-momentum tensor for the
gravitational field (see, e.g., Refs.~\cite{Babak:1999dc,Butcher:2008rf,Szabados:2009eka,Butcher:2010ja,Butcher:2012th}) 
in the current work we use the definition of the energy-momentum pseudotensor and the full four-momentum of the matter and
gravitational fields given in the classic textbook by Landau and Lifshitz \cite{Landau:1982dva}.

Our work is orginized as follows: In section~\ref{three} we specify the details of the considered EFT and calculate one-
and two-loop order contributions to the vacuum energy. In section~\ref{implications} we briefly discuss the implications
of the obtained results on the cosmological constant problem. 
We summarize in section~\ref{summary} and the appendix contains the Feynman rules and  two-loop integrals required in our calculations.

\section{Vacuum energy in an EFT of general relativity on a Minkowski background}
\label{three}

In the framework of EFT the action of matter interacting with  gravity is given by the most general effective
Lagrangian of gravitational and matter fields, which is invariant under 
general coordinate transformations and other symmetries of the Standard Model,
\begin{equation}
S = \int d^4x \sqrt{-g}\, \left\{ {\cal L}_{\rm gr}(g)
+{\cal L}_{\rm m}(g,\psi)\right\} = \int d^4x \sqrt{-g}\, \left\{ \frac{2}{\kappa^2} (R-2\Lambda)+{\cal L}_{\rm gr,ho}(g)
+{\cal L}_{\rm m}(g,\psi)\right\} = S_{\rm gr}(g)+S_{\rm m}(g,\psi) ,
\label{action}
\end{equation}
where $\kappa^2=32 \pi G$, with $G=6.70881\times 10^{-39}$ ${\rm GeV}^{-2}$  the gravitational (Newton's) constant,
 $\psi$ and $g^{\mu\nu}$ denote the matter and metric fields, respectively,  $g=\det g^{\mu\nu}$, 
$\Lambda$ is the cosmological constant and $R$ the scalar curvature. Further, ${\cal L}_{\rm gr,ho}(g)$ represents 
self-interaction terms of the gravitational field with higher orders of derivatives 
and ${\cal L}_{\rm matter}(g,\psi)$ is the effective Lagrangian of the matter fields interacting with gravity. 
Experimental evidence suggests that the contributions of ${\cal L}_{\rm gr,ho}(g)$ as well as the contributions of
non-renormalizable interactions of ${\cal L}_{\rm matter}(g,\psi)$  in physical quantities  are heavily  suppressed.
Vielbein tetrad fields have to be introduced for fermionic fields interacting with the gravitational field, however,
we refrain  from giving  details on these as later we will perform calculations with bosonic degrees of freedom only.

\medskip

The low-energy  EFT of general relativity is obtained by representing the gravitational field as the sum of the
Minkowskian background and the quantum fields \cite{tHooft:1974toh}
\begin{eqnarray}
g_{\mu\nu} &=& \eta_{\mu\nu}+\kappa h_{\mu\nu},\nonumber \\
g^ {\mu\nu} &=& \eta^{\mu\nu}-\kappa h^{\mu\nu}+\kappa^2 h^\mu_\lambda h^{\lambda\nu}-\kappa^3 h^\mu_\lambda h^{\lambda}_{\sigma} h^{\sigma\nu}+\cdots ~,
\label{gexpanded}
\end{eqnarray}
and calculating physical quantities perturbatively by applying standard QFT technique. 

\medskip

The energy-momentum tensor of the matter fields coupled to the gravitational field, $T^{\mu\nu}_{\rm m}$, and the pseudotensor
of the gravitational field, $T^{\mu\nu}_{LL}$,  are given by 
\begin{eqnarray}
T^{\mu\nu}_{\rm m} (g,\psi) & = & \frac{2}{\sqrt{-g}}\frac{\delta S_{\rm m} }{\delta g_{\mu\nu}}\,,
\label{EMTMatter}
\\
T^{\mu\nu}_{\rm gr} (g)&=& 
\frac{4}{\kappa^2} \, \Lambda\,g^{\mu\nu} +  T_{LL}^{\mu\nu}(g)\,,
\label{defTs}
\end{eqnarray}
\noindent
where $T_{LL}^{\mu\nu}(g)$ is defined via \cite{Landau:1982dva}
\begin{eqnarray}
(-g)T^{\mu\nu}_{LL} (g) &=& \frac{2}{\kappa^2} \left(\frac{1}{8} \, g^{\lambda \sigma } g^{\mu \nu } g_{\alpha \gamma}
g_{\beta\delta} \, \mathfrak{g}^{\alpha \gamma},_{\sigma }
   \, \mathfrak{g}^{\beta \delta},_\lambda -\frac{1}{4} \, g^{\mu \lambda } g^{\nu\sigma } g_{\alpha ,\gamma} g_{\beta \delta }
  \, \mathfrak{g}^{\alpha\gamma},_\sigma \, \mathfrak{g}^{\beta\delta},_\lambda -\frac{1}{4} \, g^{\lambda \sigma } g^{\mu \nu } 
   g_{\beta \alpha} g_{\gamma \delta} \, \mathfrak{g}^{\alpha \gamma},_\sigma \, \mathfrak{g}^{\beta \delta},_\lambda \right.\nonumber\\
&+& \left. \frac{1}{2}\,  g^{\mu \lambda } g^{\nu\sigma } g_{\beta \alpha} g_{\gamma \delta } \, \mathfrak{g}^{\alpha \gamma},_\sigma \, \mathfrak{g}^{\beta\delta},_\lambda 
+g^{\beta \alpha } g_{\lambda \sigma }
  \, \mathfrak{g}^{\nu \sigma},_\alpha \, \mathfrak{g}^{ \mu\lambda},_\beta +\frac{1}{2} \, g^{\mu \nu } g_{\lambda \sigma }
   \, \mathfrak{g}^{\lambda \beta},_\alpha \, \mathfrak{g}^{\alpha\sigma},_\beta \right.\nonumber\\
&-& \left.g^{\mu \lambda } g_{\sigma \beta }
   \, \mathfrak{g}^{\nu \beta},_\alpha \, \mathfrak{g}^{\sigma\alpha},_\lambda 
   -g^{\nu \lambda } g_{\sigma \beta} \, \mathfrak{g}^{\mu\beta},_\alpha \, \mathfrak{g}^{\sigma \alpha},_\lambda
   +\, \mathfrak{g}^{\lambda \sigma},_\sigma \, \mathfrak{g}^{\mu\nu},_\lambda 
   - \, \mathfrak{g}^{\text{der}(\mu \lambda},_\lambda \, \mathfrak{g}^{\nu \sigma},_\sigma \right),
   \label{LLEMT}
\end{eqnarray}
with $\mathfrak{g}^{\mu\nu}=\sqrt{-g} \, g^{\mu\nu}$ and $\mathfrak{g}^{\mu\nu},_\lambda=\partial\mathfrak{g}^{\mu\nu}/\partial x^\lambda $.

The full energy-momentum tensor $T^{\mu\nu}=T^{\mu\nu}_{\rm m} (g,\psi)+T^{\mu\nu}_{\rm gr} (g)$ defines the conserved
full four-momentum of the matter and the gravitational field as \cite{Landau:1982dva}
\begin{equation}
P^\mu= \int (-g) \, T^{\mu\nu} d S_\nu\,,
\label{EMV}
\end{equation}
where the integration is carried out over any hypersurface containing the whole three-dimensional  space. Thus,
the energy of the vacuum will be zero  if the vacuum expectation value of the energy-momentum tensor times $(-g)$ vanishes.
This quantity is given by the following path integral:
\begin{eqnarray}
\langle 0| (-g) T^{\mu\nu} |0\rangle &=& 
\int {\cal D }g\, {\cal D}\psi  \, (-g)\left[ T^{\mu\nu}_{\rm gr}(g)+ T_{\rm m}^{\mu\nu}(g,\psi) \right] \exp\left\{ i \int d^4 x \, \sqrt{-g}\,\left[ {\cal L}(g,\psi) +{\cal L}_{\rm GF}\right] \right\},
\label{VacuumE}
\end{eqnarray}
where ${\cal L}_{\rm GF}$ is the gauge fixing term and the Faddeev-Popov determinant is included in the integration measure.
The cosmological constant $\Lambda$ can be uniquely fixed by demanding that the right-hand-side of  Eq.~(\ref{VacuumE}) vanishes. 
To demonstrate how one obtains a self-consistent EFT by imposing this condition, we consider a simple model 
of a massive scalar and a massive vector fields interacting with metric tensor field. It coincides with the bosonic part of the
model with spontaneously broken  Abelian gauge symmetry considered in Ref.~\cite{Burns:2014bva} 
taken in unitary gauge for the Abelian gauge symmetry. The action of the matter part of the model is given by
\begin{equation}
S_{\rm m} = \int d^4x \sqrt{-g}\, \left\{ -\frac{1}{4} g^{\mu\rho}g^{\nu\sigma} F_{\mu\nu}F_{\rho\sigma} +\frac{M^2}{2}\, g^{\mu\nu} A_\mu A_\nu  +\frac{g^{\mu\nu}}{2}\,\partial_\mu H \partial_\nu H -\frac{m^2}{2}\,H^2+{\cal L}_{\rm MI}\right\},
\label{MAction}
\end{equation}
where $F_{\mu\nu} = \partial_\mu A_\nu -\partial_\nu A_\mu$, $A_\mu$ is vector field, $H$ the scalar field and ${\cal L}_{\rm MI}$
denotes the  interactions of matter fields, the specific form of which is not important for the current work as we will not 
include them in our calculations. The energy-momentum tensor corresponding to Eq.~(\ref{MAction}) has the form
\begin{eqnarray}
T_m^{\mu\nu} & = &  - g^{\mu\alpha}g^{\nu\rho}g^{\beta\sigma} F_{\alpha\beta}F_{\rho\sigma} + M^2\, g^{\mu\alpha}g^{\nu\beta} A_\alpha A_\beta +\partial_\mu H \partial_\nu H \nonumber\\
&-& g^{\mu\nu} \left\{ -\frac{1}{4} g^{\alpha\rho}g^{\beta\sigma} F_{\alpha\beta}F_{\rho\sigma} +\frac{M^2}{2}\, g^{\alpha\beta} A_\alpha A_\beta  +\frac{g^{\alpha\beta}}{2}\,\partial_\alpha H \partial_\beta H -\frac{m^2}{2}\,H^2\right\} +T_{\rm MI}^{\mu\nu} ,
\label{MEMT}
\end{eqnarray}
where $T_{\rm MI}^{\mu\nu}$ corresponds to ${\cal L}_{\rm MI}$. 

By adding the following gauge fixing term to the effective Lagrangian
\begin{equation}
{\cal L}_{\rm GF}= \xi \left( \partial_\nu h^{\mu\nu}-\frac{1}{2} \partial^\mu h^\nu_\nu \right) \left( \partial^\beta h_{\mu\beta}-\frac{1}{2} \partial_\mu h^\alpha_\alpha \right),
\label{GFT}
\end{equation}
where $\xi$ is the gauge parameter, we obtain the Feynman rules specified in the appendix.

\begin{figure}[t]
\begin{center}
 \includegraphics[height=3.00cm]{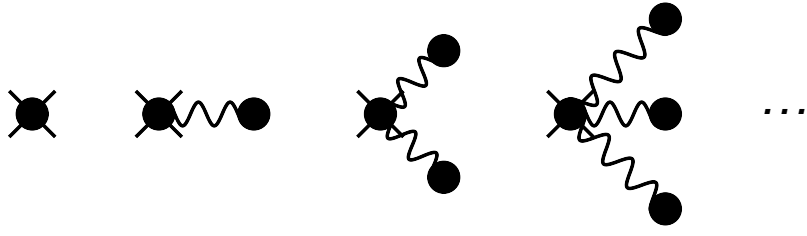}   
 \caption{Diagrams contributing to the vacuum expectation value of the energy-momentum pseudotensor times $(-g)$ at tree
   order. Filled circles corresponds to the cosmological constant term. 
  The cross stands for the energy-momentum pseudotensor times $(-g)$, and the wiggly line represents the graviton.} \label{EMTTree}
\end{center}
\end{figure}

\medskip

For the vacuum expectation value of the full energy-momentum pseudotensor times $(-g)$ at tree order we obtain an
infinite number of diagrams shown in Fig.~\ref{EMTTree}. All these contributions vanish if we take the cosmological
constant vanishing at tree order.   That is, we represent $\Lambda$  as 
\begin{equation}
\Lambda = \sum_{i=0}^\infty \hbar^i \Lambda _i \,,
\label{CCexpanded}
\end{equation}
and take $\Lambda_0=0$.
Notice that this also removes the graviton mass from the propagator at tree order.

\begin{figure}[t]
  \begin{center}
    \includegraphics[height=2.0cm]{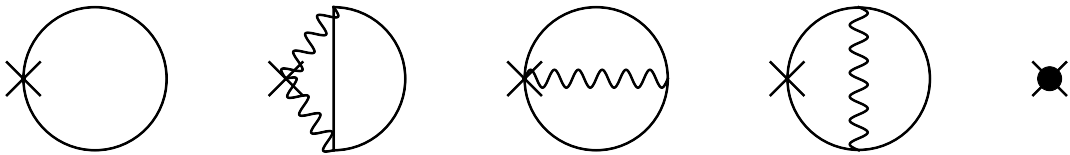}  
    \caption{Diagrams contributing to the vacuum expectation value of the energy-momentum  pseudotensor times $(-g)$.
     Filled circle corresponds to the cosmological constant term. 
The cross stands for the energy-momentum pseudotensor times $(-g)$, wiggly and solid lines represent the graviton and the scalar (vector), respectively.}
\label{EMTLoop}
\end{center}
\end{figure}

\medskip

\begin{figure}[t]
\begin{center}
 \includegraphics[height=2.25cm]{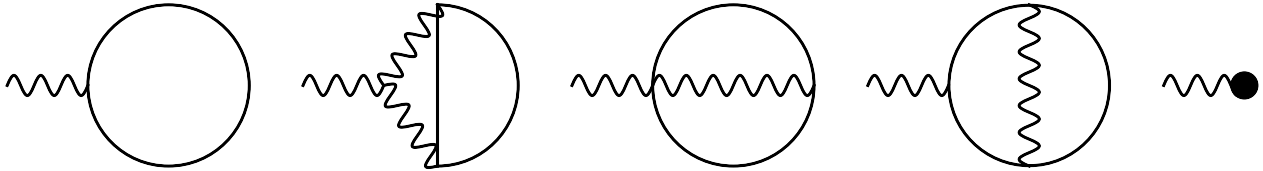}   
 \caption{Diagrams contributing to the graviton tadpole. The filled circle corresponds to the cosmological constant term.
   Wiggly and solid lines represent the graviton and the scalar (vector) fields, respectively. }
\label{GrTP}
\end{center}
\end{figure}

Next, using the Feynman rules given in the appendix, we calculated the one-loop contributions to the vacuum expectation value of
the full energy-momentum pseudotensor  times $(-g)$  shown in Fig.~\ref{EMTLoop},  
and by demanding that $\Lambda_1$ cancels this contribution we obtain (in the calculations of the loop diagrams below we used the program
FeynCalc \cite{Mertig:1990an,Shtabovenko:2016sxi})
\begin{equation}
\Lambda_1 = -\frac{ \kappa ^2 \Gamma
   \left(1-\frac{d}{2}\right) \left(m^d+(d-1)
   M^d\right)}{ 2^{d+6} \pi ^{\frac{d}{2}+4} \, d}\,.
\end{equation}
It is a trivial consequence of Eq.~(\ref{EMTMatter}) that the same value of $\Lambda_1$ cancels the one-loop contribution
to the vacuum expectation value of the graviton field $h_{\mu\nu}$, shown in Fig.~\ref{GrTP}, and 
consequently the graviton self-energy at zero momentum, i.e. graviton mass, as a result of a Ward identity \cite{Burns:2014bva}.
The first non-trivial result is obtained at two-loop order by calculating the diagrams contributing
to the vacuum expectation value of the full energy-momentum pseudotensor times $(-g)$ shown in Fig.~\ref{EMTLoop}. 
We also calculated the two-loop contributions to the vacuum expectation value of the gravitational field shown
in Fig.~\ref{GrTP}  and verified that the same value of $\Lambda_2$ cancels both quantities. 
The obtained result reads:
\begin{equation}
\Lambda_2 = -\frac{ d (d+1)  \kappa ^4 M^{2
   d-2} \csc \left(\frac{\pi  d}{2}\right) \Gamma
   \left(1-\frac{d}{2}\right)}{2^{2 (d+3)} \pi ^{d-1}\Gamma \left(\frac{d}{2}\right)}.
\label{Lambda2}
\end{equation}
While it is a trivial consequence of Eq.~(\ref{EMTMatter}) that the fourth diagrams in both figures \ref{EMTLoop} and \ref{GrTP}
give equal contributions in $\Lambda$ it is only the sum of the corresponding second and third diagrams that lead to
identical expressions.
To check the obtained results we also calculated two-loop contributions to the graviton self-energy at zero momentum
and verified that the same value of $\Lambda_2$ cancels the two-loop order contribution to the graviton 
mass in agreement with the Ward identity \cite{Burns:2014bva} (we do not give the expressions of the corresponding
Feynman rules due to their huge size).  
While we expect that an analogous result holds to all orders we are not able to give a general argument supporting it.

\section{Implication on the cosmological constant problem}
\label{implications}

\begin{figure}[t]
\begin{center}
\epsfig{file=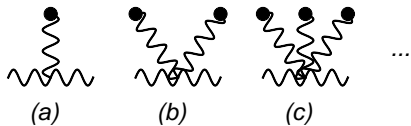,scale=0.5}
\caption{Tree order tadpole diagrams contributing to the graviton self-energy.}
\label{GrSE}
\end{center}
\vspace{-5mm}
\end{figure}

It follows from the result of the previous section that unless the cosmological constant is chosen such that the energy
of the vacuum is exactly zero, it cannot remove the graviton mass and the graviton tadpole  order-by-order in
perturbation theory and consequently a non-perturbative treatment of the cosmological constant term is mandatory. This is
because for all physical processes there appear diagrams like ones shown in Fig.~\ref{GrSE} where the massless graviton
propagator carries  vanishing momentum and therefore $1/0$ 
singularities occur (this does not happen only if the tadpole vanishes order-by-order in the loop expansion). 

The cosmological constant problem is often described as vacuum having tiny non-zero energy density. Due to
this loose language one might think that the condition of vanishing vacuum energy {\it a priory} excludes the solution
of the cosmological constant problem. A closer look reveals that exactly the opposite might be the case. Indeed, due to
the condition imposed on the cosmological constant term of the effective Lagrangian the effective action calculated
on the Minkowski background metric with vanishing background matter fields does not contain an effective cosmological
constant term contributing to Einsten's equations. However, for our universe the corresponding effective action has to
be calculated in the presence of non-trivial background fields. The cosmological constant term of the effective Lagrangian 
exactly cancelling the loop contributions in a trivial background leads to a uniquely fixed effective cosmological constant
contributing in the Einstein's equation also in the presence of a non-trivial background. While for weak backgrounds we expect
large cancellations leaving us with a tiny effective cosmological constant, a quantitative investigation of this estimation
is a subject of a separate publication. To make this more precise, the background relevant for cosmology is not flat Minkowski,
therefore the fixed cosmological constant term will not exactly cancel the quantum corrections to the effective cosmological
constant, but rather only approximately, leaving a small finite piece. This remains to be calculated.

 \section{Summary and discussion}
\label{summary}

Consistency conditions of the perturbative EFT of general relativity in flat Minkowski background uniquely fix
the cosmological constant term as a function of all other parameters of the theory \cite{Burns:2014bva}. 
This follows from the requirement of the presence of a massless graviton, instead of a massive spin-two ghost,
in the spectrum of the theory. 
Notice that it is not possible to take into account perturbatively any other value of the cosmological constant
term within an EFT on the flat Minkowski background. This is because of the $1/0$ singularities 
in the Feynman diagrams with tadpole contributions, see, e.g., Fig.~\ref{GrSE}. 
 
In our opinion if there is any fundamental reason for choosing a fixed value of the cosmological constant then it
must be the condition of vanishing of the vacuum energy. 
It is often argued that vacuum has non-zero energy due to  quantum fluctuations. 
A classical example is given by quantum oscillator. It is well-known that the ground state energy of a
quantum oscillator is $\hbar \omega/2$, where $\omega$ is the angular frequency.  
A closer look reveals, however, that this expression is a result of an assumption. In particular, if we share
the point of view that the real world is described by a quantum theory and the classical theory is only an
approximation to it,  then it is not possible to uniquely reproduce the quantum Hamiltonian of an oscillator by
quantising the classical one. This non-uniqueness is of course well-known and is manifested in the problem of
operator ordering. Indeed, by  adding  a vanishing term $\sim (p q-q p)$ to the classical Hamiltonian of the oscillator
and quantizing it we obtain a quantum Hamiltonian with an arbitarary constant term and hence an arbitrary vacuum energy.  
Starting from the classical theory there is no way to tell which value of the vacuum energy is more ``fundamental".
Notice that the argument of the Casimir effect being a proof of the non-vanishing vacuum energy is not convincing
either, see, e.g., Refs.~\cite{Jaffe:2005vp,Nikolic:2016kkp}.

In the framework of low-energy EFT of general relativity coupled to the fields of the Standard Model imposing a
condition of vanishing vacuum energy uniquely fixes the cosmological constant term as a function of other
parameters of the effective Lagrangian. 
We expect that this leads to a self-consistent perturbative EFT defined on the Minkowsky background, i.e.
to a massless graviton in the spectrum and the vanishing graviton tadpole. We were not able to give a general
argument supporting our claim.
Instead we calculated the vacuum expectation value of the full four-momentum of matter and gravitational
fields at two-loop order in a simplified version of the Abelian model with spontaneous symmetry breaking
considered in Ref.~\cite{Burns:2014bva}. 
While at one-loop order the condition of vanishing vacuum energy automatically leads to the conditions of
Ref.~\cite{Burns:2014bva}, at two-loop order 
the same agreement of two conditions appears as a result of a non-trivial cancellation between different diagrams.
We notice here that there does not exist a commonly accepted expression of the energy-momentum tensor for the
gravitational field (see, e.g., Refs.~\cite{Babak:1999dc,Butcher:2008rf,Szabados:2009eka,Butcher:2010ja,Butcher:2012th}). 
In the current work we used the definition of the energy-momentum pseudotensor and the full
four-momentum of the matter and gravitational fields given in
the classic textbook by Landau and Lifshitz \cite{Landau:1982dva}. 

Within a self-consistent EFT all physical quantities should be finite after renormalizing (an infinite number of)
parameters of the effective Lagrangian. 
Therefore it is mandatory that the uniquely fixed value of the cosmological constant term, which defines the
perturbative EFT of the Standard Model coupled to gravitons on Minkowski flat background leads to a finite expression of the 
energy of the vacuum to all orders in loop expansion. 
Based on the two-loop order result of the current work we expect that this finite value is actually zero.
Turning the argument around we expect that by demanding that the vacuum energy should be vanishing to all orders 
we obtain a self-consistent perturbative low-energy EFT of matter and gravitational fields on flat Minkowski background.

Relegating calculations and detailed discussion to a future work, we briefly comment on the implications of our results
for the cosmological constant problem. In particular, we expect that the cosmological constant term of the effective Lagrangian 
exactly cancelling the loop contributions in flat background is very likely to cancel the bulk of such contributions also
in the presence of a non-trivial background, relevant for our universe, 
thus leaving with a tiny effective cosmological constant contributing to Einstein's equations.

\medskip

\acknowledgments

We thank Dalibor Djukanovic for helpful comments on the manuscript. 
The work of JG was supported in part by BMBF (Grant No. 05P18PCFP1), and by the
Georgian Shota Rustaveli National Science Foundation (Grant No. FR17-354).
The work of UGM was supported in part by provided by Deutsche
Forschungsgemeinschaft (DFG)  through funds provided to the Sino-German CRC~110
``Symmetries and the Emergence of Structure in QCD" (Grant No. TRR110),  by
the Chinese Academy of Sciences (CAS) through a President's International Fellowship
Initiative (PIFI) (Grant No. 2018DM0034) and by the VolkswagenStiftung (Grant No. 93562).

\section*{Appendix}

Below we give Feynman rules used in the calculation of the vacuum expectation values of the graviton field
and the energy-momentum tensor. 

\medskip

Propagators:

\begin{itemize}

\item
Scalar propagator with  momentum $p$:
\begin{equation}
 \frac{ i }{ p^2-m^2+i
   \epsilon }\,.
\label{sPr}
\end{equation}

\item
Vector boson propagator with Lorentz indices $\mu$, $\nu$ and momentum $p$:
\begin{equation}
-\frac{i \left( g^{\mu \nu} -p^\mu p^\nu/M^2\right)}{p^2-M^2+i \epsilon}  \,.
\label{vPr}
\end{equation}

\item
Graviton propagator in $D$ dimensions with Lorentz indices $(\mu,\nu)$, $(\alpha,\beta)$ and momentum $p$:
\begin{equation}
\frac{i}{2} \, \frac{ g^{\lambda \nu } g^{\mu \sigma }+g^{\lambda \mu } g^{\nu
   \sigma } - \frac{2 g^{\lambda \sigma } g^{\mu \nu
   }}{D-2} }{p^2+i \epsilon } - \frac{i \, \xi}{2} \frac{ 
    p^{\nu } \left(p^{\sigma } g^{\lambda \mu }+p^{\lambda }
   g^{\mu \sigma }\right)+p^{\mu } \left(p^{\sigma } g^{\lambda \nu
   }+p^{\lambda } g^{\nu \sigma }\right) }{ \left(p^2+i
   \epsilon \right)^2} \,.
\label{gPr}
\end{equation}

\medskip

Vertices (all momenta in all vertices are incoming):

\item 
Graviton with indices $(\mu,\nu)$:
\begin{equation}
-\frac{2 i \Lambda  g^{\mu \nu }}{\kappa } ;
\label{h}
\end{equation}

\item
Graviton with indices $(\mu,\nu)$ and $(\alpha,\beta)$:
\begin{equation}
i \Lambda  \left(g^{\alpha \nu } g^{\beta \mu }+g^{\alpha \mu }
   g^{\beta \nu }-g^{\alpha \beta } g^{\mu \nu }\right) ;
\label{h}
\end{equation}

\item
Graviton with indices $(\mu,\nu)$ - scalars with momenta $p_1$ and $p_2$:
\begin{equation}
\frac{1}{2} i \kappa  \left(-g^{\mu \nu } \left(m^2+p_1\cdot
   p_2\right)+p_2^{\mu } p_1^{\nu }+p_1^{\mu } p_2^{\nu }\right) ;
\label{hSS}
\end{equation}

\item
Gravitons with indices $(\mu,\nu)$ and $(\alpha,\beta)$ - scalars with momenta $p_1$ and $p_2$:
\begin{eqnarray}
&& -\frac{1}{4} i \kappa ^2 \left(-m^2 g^{\alpha \nu }
   g^{\beta \mu }-m^2 g^{\alpha \mu } g^{\beta \nu }+m^2 g^{\alpha
   \beta } g^{\mu \nu }+p_1^{\beta } p_2^{\nu } g^{\alpha \mu
   }+p_1^{\beta } p_2^{\mu } g^{\alpha \nu }+p_1^{\alpha } p_2^{\nu
   } g^{\beta \mu } 
   \right.  \nonumber\\
   && \left.
   +p_1^{\nu } \left(-p_2^{\mu } g^{\alpha \beta
   }+p_2^{\beta } g^{\alpha \mu }+p_2^{\alpha } g^{\beta \mu
   }\right)+p_1^{\alpha } p_2^{\mu } g^{\beta \nu }+p_1^{\mu }
   \left(-p_2^{\nu } g^{\alpha \beta }+p_2^{\beta } g^{\alpha \nu
   }+p_2^{\alpha } g^{\beta \nu }\right)-p_2^{\alpha } p_1^{\beta }
   g^{\mu \nu } 
   \right.  \nonumber\\
   && \left.
   -p_1^{\alpha } p_2^{\beta } g^{\mu \nu }-p_1\cdot
   p_2 \left(g^{\alpha \nu } g^{\beta \mu }+g^{\alpha \mu }
   g^{\beta \nu }-g^{\alpha \beta } g^{\mu \nu }\right)\right) ;
\label{hhSS}
\end{eqnarray}

\item
Graviton with indices $(\mu,\nu)$ - vector bosons with (Lorentz index, momentum) combinations $(\lambda,p_1)$  and $(\sigma, p_2)$:
\begin{eqnarray}
&& -\frac{i}{2}  \kappa  \left( - M^2 g^{\lambda \sigma }
   g^{\mu \nu }+M^2 g^{\lambda \nu } g^{\mu \sigma }+M^2 g^{\lambda
   \mu } g^{\nu \sigma }+p_1^{\mu } p_2^{\nu } g^{\lambda \sigma
   }-p_1^{\sigma } \left(p_2^{\nu } g^{\lambda \mu }+p_2^{\mu }
   g^{\lambda \nu }-p_2^{\lambda } g^{\mu \nu }\right) 
   \right.  \nonumber\\
   && \left.
   +p_1^{\nu }
   \left(p_2^{\mu } g^{\lambda \sigma }-p_2^{\lambda } g^{\mu
   \sigma }\right)-p_2^{\lambda } p_1^{\mu } g^{\nu \sigma
   }-p_1\cdot p_2 g^{\lambda \sigma } g^{\mu \nu }+p_1\cdot p_2
   g^{\lambda \nu } g^{\mu \sigma }+p_1\cdot p_2 g^{\lambda \mu }
   g^{\nu \sigma }\right);
\label{hVV}
\end{eqnarray}

\item
  Gravitons with indices $(\mu,\nu)$ and $(\alpha,\beta)$ - vector bosons with (Lorentz index, momentum)
  combinations $(\lambda,p_1)$  and $(\sigma, p_2)$:
\begin{eqnarray}
&& 
-\frac{i}{4}  \kappa ^2 \left(-g^{\alpha \sigma } g^{\beta \nu }
   g^{\lambda \mu } M^2-g^{\alpha \nu } g^{\beta \sigma }
   g^{\lambda \mu } M^2-g^{\alpha \sigma } g^{\beta \mu }
   g^{\lambda \nu } M^2-g^{\alpha \mu } g^{\beta \sigma }
   g^{\lambda \nu } M^2+g^{\alpha \nu } g^{\beta \mu } g^{\lambda
   \sigma } M^2
   \right.  \nonumber\\
   && \left.
   +g^{\alpha \mu } g^{\beta \nu } g^{\lambda \sigma }
   M^2+g^{\alpha \sigma } g^{\beta \lambda } g^{\mu \nu }
   M^2+g^{\alpha \lambda } g^{\beta \sigma } g^{\mu \nu }
   M^2-g^{\alpha \beta } g^{\lambda \sigma } g^{\mu \nu }
   M^2-g^{\alpha \nu } g^{\beta \lambda } g^{\mu \sigma }
   M^2-g^{\alpha \lambda } g^{\beta \nu } g^{\mu \sigma }
   M^2
   \right.  \nonumber\\
   && \left.
   +g^{\alpha \beta } g^{\lambda \nu } g^{\mu \sigma }
   M^2-g^{\alpha \mu } g^{\beta \lambda } g^{\nu \sigma }
   M^2-g^{\alpha \lambda } g^{\beta \mu } g^{\nu \sigma }
   M^2+g^{\alpha \beta } g^{\lambda \mu } g^{\nu \sigma }
   M^2-p_1^{\mu } p_2^{\nu } g^{\alpha \sigma } g^{\beta \lambda
   }+p_1^{\mu } p_2^{\lambda } g^{\alpha \sigma } g^{\beta \nu
   } 
   \right.  \nonumber\\
   && \left.
   -p_1^{\mu } p_2^{\nu } g^{\alpha \lambda } g^{\beta \sigma
   }+p_1^{\mu } p_2^{\lambda } g^{\alpha \nu } g^{\beta \sigma
   }+p_1^{\beta } p_2^{\nu } g^{\alpha \sigma } g^{\lambda \mu
   }+p_1^{\alpha } p_2^{\nu } g^{\beta \sigma } g^{\lambda \mu
   }+p_1^{\beta } p_2^{\mu } g^{\alpha \sigma } g^{\lambda \nu
   }+p_1^{\alpha } p_2^{\mu } g^{\beta \sigma } g^{\lambda \nu
   }+p_1^{\mu } p_2^{\nu } g^{\alpha \beta } g^{\lambda \sigma
   }
   \right.  \nonumber\\
   && \left.
   -p_1^{\beta } p_2^{\nu } g^{\alpha \mu } g^{\lambda \sigma
   }-p_1^{\mu } p_2^{\beta } g^{\alpha \nu } g^{\lambda \sigma
   }-p_1^{\beta } p_2^{\mu } g^{\alpha \nu } g^{\lambda \sigma
   }-p_1^{\alpha } p_2^{\nu } g^{\beta \mu } g^{\lambda \sigma
   }-p_1^{\mu } p_2^{\alpha } g^{\beta \nu } g^{\lambda \sigma
   }-p_1^{\alpha } p_2^{\mu } g^{\beta \nu } g^{\lambda \sigma
   }-p_1^{\beta } p_2^{\lambda } g^{\alpha \sigma } g^{\mu \nu
   }
   \right.  \nonumber\\
   && \left.
   -p_1^{\alpha } p_2^{\lambda } g^{\beta \sigma } g^{\mu \nu
   }+p_1^{\beta } p_2^{\alpha } g^{\lambda \sigma } g^{\mu \nu
   }+p_1^{\alpha } p_2^{\beta } g^{\lambda \sigma } g^{\mu \nu
   }+p_1^{\sigma } \left(p_2^{\mu } g^{\alpha \nu } g^{\beta
   \lambda }-p_2^{\alpha } g^{\mu \nu } g^{\beta \lambda
   }
   -p_2^{\lambda } g^{\alpha \nu } g^{\beta \mu }+p_2^{\mu }
   g^{\alpha \lambda } g^{\beta \nu }
      \right. \right.  \nonumber\\
   && \left. \left.
   -p_2^{\lambda } g^{\alpha \mu
   } g^{\beta \nu }+p_2^{\alpha } g^{\beta \nu } g^{\lambda \mu
   }+p_2^{\nu } \left(g^{\alpha \mu } g^{\beta \lambda }+g^{\alpha
   \lambda } g^{\beta \mu }-g^{\alpha \beta } g^{\lambda \mu
   }\right)-p_2^{\mu } g^{\alpha \beta } g^{\lambda \nu
   }+p_2^{\alpha } g^{\beta \mu } g^{\lambda \nu }+p_2^{\lambda }
   g^{\alpha \beta } g^{\mu \nu }
      \right. \right.  \nonumber\\
   && \left. \left.
   +p_2^{\beta } \left(g^{\alpha \nu
   } g^{\lambda \mu }+g^{\alpha \mu } g^{\lambda \nu }-g^{\alpha
   \lambda } g^{\mu \nu }\right)\right)+p_1^{\beta } p_2^{\lambda }
   g^{\alpha \nu } g^{\mu \sigma }+p_1^{\alpha } p_2^{\lambda }
   g^{\beta \nu } g^{\mu \sigma }-p_1^{\beta } p_2^{\alpha }
   g^{\lambda \nu } g^{\mu \sigma }-p_1^{\alpha } p_2^{\beta }
   g^{\lambda \nu } g^{\mu \sigma }
      \right.   \nonumber\\
   && \left. 
    +p_1^{\nu } \left(-p_2^{\mu }
   g^{\alpha \sigma } g^{\beta \lambda }+p_2^{\alpha } g^{\mu
   \sigma } g^{\beta \lambda }-p_2^{\mu } g^{\alpha \lambda }
   g^{\beta \sigma }+p_2^{\mu } g^{\alpha \beta } g^{\lambda \sigma
   }-p_2^{\alpha } g^{\beta \mu } g^{\lambda \sigma }+p_2^{\lambda
   } \left(g^{\alpha \sigma } g^{\beta \mu }+g^{\alpha \mu }
   g^{\beta \sigma }-g^{\alpha \beta } g^{\mu \sigma
   }\right)
     \right. \right.  \nonumber\\
   && \left. \left.  
  +p_2^{\beta } \left(g^{\alpha \lambda } g^{\mu \sigma
   }-g^{\alpha \mu } g^{\lambda \sigma }\right)\right)-p_1^{\mu }
   p_2^{\lambda } g^{\alpha \beta } g^{\nu \sigma }+p_1^{\mu }
   p_2^{\beta } g^{\alpha \lambda } g^{\nu \sigma }+p_1^{\beta }
   p_2^{\lambda } g^{\alpha \mu } g^{\nu \sigma }+p_1^{\mu }
   p_2^{\alpha } g^{\beta \lambda } g^{\nu \sigma }+p_1^{\alpha }
   p_2^{\lambda } g^{\beta \mu } g^{\nu \sigma }
      \right.   \nonumber\\
   && \left. 
   -p_1^{\beta }
   p_2^{\alpha } g^{\lambda \mu } g^{\nu \sigma }-p_1^{\alpha }
   p_2^{\beta } g^{\lambda \mu } g^{\nu \sigma }-g^{\alpha \sigma }
   g^{\beta \nu } g^{\lambda \mu } p_1\cdot p_2-g^{\alpha \nu }
   g^{\beta \sigma } g^{\lambda \mu } p_1\cdot p_2-g^{\alpha \sigma
   } g^{\beta \mu } g^{\lambda \nu } p_1\cdot p_2 
      \right. \nonumber\\
   && \left. 
   -g^{\alpha \mu }
   g^{\beta \sigma } g^{\lambda \nu } p_1\cdot p_2+g^{\alpha \nu }
   g^{\beta \mu } g^{\lambda \sigma } p_1\cdot p_2+g^{\alpha \mu }
   g^{\beta \nu } g^{\lambda \sigma } p_1\cdot p_2+g^{\alpha \sigma
   } g^{\beta \lambda } g^{\mu \nu } p_1\cdot p_2+g^{\alpha \lambda
   } g^{\beta \sigma } g^{\mu \nu } p_1\cdot p_2
   \right. \nonumber\\
   && \left. 
   -g^{\alpha \beta }
   g^{\lambda \sigma } g^{\mu \nu } p_1\cdot p_2-g^{\alpha \nu }
   g^{\beta \lambda } g^{\mu \sigma } p_1\cdot p_2-g^{\alpha
   \lambda } g^{\beta \nu } g^{\mu \sigma } p_1\cdot p_2+g^{\alpha
   \beta } g^{\lambda \nu } g^{\mu \sigma } p_1\cdot p_2-g^{\alpha
   \mu } g^{\beta \lambda } g^{\nu \sigma } p_1\cdot p_2
   \right. \nonumber\\
   && \left. 
   -g^{\alpha
   \lambda } g^{\beta \mu } g^{\nu \sigma } p_1\cdot p_2+g^{\alpha
   \beta } g^{\lambda \mu } g^{\nu \sigma } p_1\cdot p_2\right);
\label{hhVV}
\end{eqnarray}

\item
  Energy-momentum tensor with indices $(\mu,\nu)$ - gravitons with (Lorentz indices, momentum)
  combinations $(\lambda,\sigma, p_1)$  and $(\alpha,\beta, p_2)$:
\begin{eqnarray}
&&
\frac{1}{8} \left[\text{hhh}\left(\left\{\mu ,\nu
   ,p_1\right\},\left\{\alpha ,\beta ,p_2\right\},\left\{\lambda
   ,\sigma ,p_3\right\}\right)+\text{hhh}\left(\left\{\mu ,\nu
   ,p_1\right\},\left\{\alpha ,\beta ,p_2\right\},\left\{\sigma
   ,\lambda ,p_3\right\}\right) \right. \nonumber\\
&&   
\left.
   +\text{hhh}\left(\left\{\mu ,\nu
   ,p_1\right\},\left\{\beta ,\alpha ,p_2\right\},\left\{\lambda
   ,\sigma ,p_3\right\}\right)+\text{hhh}\left(\left\{\mu ,\nu
   ,p_1\right\},\left\{\beta ,\alpha ,p_2\right\},\left\{\sigma
   ,\lambda ,p_3\right\}\right)
   \right. \nonumber\\
&&   
\left.
   +\text{hhh}\left(\left\{\nu ,\mu
   ,p_1\right\},\left\{\alpha ,\beta ,p_2\right\},\left\{\lambda
   ,\sigma ,p_3\right\}\right)+\text{hhh}\left(\left\{\nu ,\mu
   ,p_1\right\},\left\{\alpha ,\beta ,p_2\right\},\left\{\sigma
   ,\lambda ,p_3\right\}\right)
   \right. \nonumber\\
&&   
\left.
   +\text{hhh}\left(\left\{\nu ,\mu
   ,p_1\right\},\left\{\beta ,\alpha ,p_2\right\},\left\{\lambda
   ,\sigma ,p_3\right\}\right)+\text{hhh}\left(\left\{\nu ,\mu
   ,p_1\right\},\left\{\beta ,\alpha ,p_2\right\},\left\{\sigma
   ,\lambda ,p_3\right\}\right)\right],
\end{eqnarray}
where
\begin{eqnarray}
&&
\text{hhh}\left(\left\{\mu ,\nu
   ,p_1\right\},\left\{\alpha ,\beta ,p_2\right\},\left\{\lambda
   ,\sigma ,p_3\right\}\right)=-\frac{1}{4} i \kappa  \left(p_1^{\alpha } p_2^{\beta } g^{\lambda
   \sigma } g^{\mu \nu }+p_2^{\beta } p_3^{\alpha } g^{\lambda
   \sigma } g^{\mu \nu }+2 \Lambda  g^{\alpha \beta } g^{\lambda
   \sigma } g^{\mu \nu }
   \right. \nonumber\\
&&   
\left.
   +2 g^{\alpha \beta } g^{\lambda \sigma }
   \left(p_1\cdot p_2+p_1\cdot p_3+p_2\cdot p_3\right) g^{\mu \nu
   }+2 g^{\alpha \beta } g^{\lambda \sigma }
   \left(p_1^2+p_2^2+p_3^2\right) g^{\mu \nu }
   \right. \nonumber\\
&&   
\left.
   +g^{\alpha \beta }
   \left(p_1^{\nu } \left(p_2^{\mu }+p_3^{\mu }\right) g^{\lambda
   \sigma }+\left(p_1^{\lambda }+p_2^{\lambda }\right) p_3^{\sigma
   } g^{\mu \nu }\right)+4 \left(p_1^{\sigma } p_3^{\nu } g^{\alpha
   \beta } g^{\lambda \mu }+p_1^{\beta } p_2^{\nu } g^{\alpha \mu }
   g^{\lambda \sigma }+p_2^{\sigma } p_3^{\beta } g^{\alpha \lambda
   } g^{\mu \nu }\right)
   \right. \nonumber\\
&&   
\left.
   +4 \left(\left(p_1^{\sigma } p_2^{\nu
   }+p_2^{\sigma } p_3^{\nu }\right) g^{\alpha \beta } g^{\lambda
   \mu }+\left(p_2^{\nu } p_3^{\beta }+p_1^{\beta } p_3^{\nu
   }\right) g^{\alpha \mu } g^{\lambda \sigma }+\left(p_1^{\beta }
   p_2^{\sigma }+p_1^{\sigma } p_3^{\beta }\right) g^{\alpha
   \lambda } g^{\mu \nu }\right)
   \right. \nonumber\\
&&   
\left.
   -2 \left(\left(p_1^{\nu }
   p_2^{\sigma }+p_2^{\nu } p_3^{\sigma }\right) g^{\alpha \beta }
   g^{\lambda \mu }+\left(p_1^{\nu } p_3^{\beta }+p_2^{\beta }
   p_3^{\nu }\right) g^{\alpha \mu } g^{\lambda \sigma
   }+\left(p_1^{\sigma } p_2^{\beta }+p_1^{\beta } p_3^{\sigma
   }\right) g^{\alpha \lambda } g^{\mu \nu }\right)
   \right. \nonumber\\
&&   
\left.
   +2
   \left(\left(p_1^{\nu } p_1^{\sigma }+p_3^{\nu } p_3^{\sigma
   }\right) g^{\alpha \beta } g^{\lambda \mu }+\left(p_1^{\beta }
   p_1^{\nu }+p_2^{\beta } p_2^{\nu }\right) g^{\alpha \mu }
   g^{\lambda \sigma }+\left(p_2^{\beta } p_2^{\sigma }+p_3^{\beta
   } p_3^{\sigma }\right) g^{\alpha \lambda } g^{\mu \nu }\right)
   \right. \nonumber\\
&&   
\left.
   -2
   \left(p_1^{\mu } p_1^{\nu } g^{\alpha \beta } g^{\lambda \sigma
   }+\left(p_3^{\lambda } p_3^{\sigma } g^{\alpha \beta
   }+p_2^{\alpha } p_2^{\beta } g^{\lambda \sigma }\right) g^{\mu
   \nu }\right)-2 \left(p_1^{\beta } p_3^{\alpha } g^{\lambda
   \sigma } g^{\mu \nu }+p_1^{\alpha } p_3^{\beta } g^{\lambda
   \sigma } g^{\mu \nu }
   \right. \right. \nonumber\\
&&   
\left. \left.
   +g^{\alpha \beta } \left(\left(p_2^{\nu }
   p_3^{\mu }+p_2^{\mu } p_3^{\nu }\right) g^{\lambda \sigma
   }+\left(p_1^{\sigma } p_2^{\lambda }+p_1^{\lambda } p_2^{\sigma
   }\right) g^{\mu \nu }\right)\right)-4 \left(p_1^{\alpha }
   p_1^{\beta } g^{\lambda \sigma } g^{\mu \nu }+p_3^{\alpha }
   p_3^{\beta } g^{\lambda \sigma } g^{\mu \nu } 
   \right. \right. \nonumber\\
&&   
\left. \left.
    +g^{\alpha \beta }
   \left(\left(p_2^{\mu } p_2^{\nu }+p_3^{\mu } p_3^{\nu }\right)
   g^{\lambda \sigma }+\left(p_1^{\lambda } p_1^{\sigma
   }+p_2^{\lambda } p_2^{\sigma }\right) g^{\mu \nu
   }\right)\right)-5 \left(p_2^{\alpha } \left(p_1^{\beta
   }+p_3^{\beta }\right) g^{\lambda \sigma } g^{\mu \nu }
    +g^{\alpha
   \beta } \left(p_1^{\mu } \left(p_2^{\nu }+p_3^{\nu }\right)
   g^{\lambda \sigma } 
    \right. \right. \right. \nonumber\\
&&   
\left. \left. \left.
   +\left(p_1^{\sigma }+p_2^{\sigma }\right)
   p_3^{\lambda } g^{\mu \nu }\right)\right)
   +2 \left(p_3^{\mu }
   p_3^{\sigma } g^{\alpha \beta } g^{\lambda \nu }+p_1^{\alpha }
   p_1^{\nu } g^{\beta \mu } g^{\lambda \sigma }+p_3^{\alpha }
   p_3^{\sigma } g^{\beta \lambda } g^{\mu \nu }+p_2^{\beta }
   \left(p_2^{\mu } g^{\alpha \nu } g^{\lambda \sigma
   }+p_2^{\lambda } g^{\alpha \sigma } g^{\mu \nu
   }\right)
    \right. \right. \nonumber\\
&&   
\left. \left.
   +p_1^{\lambda } p_1^{\nu } g^{\alpha \beta } g^{\mu
   \sigma }\right)+2 \left(p_2^{\sigma } p_3^{\mu } g^{\alpha \beta
   } g^{\lambda \nu }+p_2^{\mu } p_3^{\beta } g^{\alpha \nu }
   g^{\lambda \sigma }+p_1^{\alpha } p_3^{\nu } g^{\beta \mu }
   g^{\lambda \sigma }+p_1^{\beta } p_2^{\lambda } g^{\alpha \sigma
   } g^{\mu \nu }+p_1^{\sigma } p_3^{\alpha } g^{\beta \lambda }
   g^{\mu \nu }+p_1^{\lambda } p_2^{\nu } g^{\alpha \beta } g^{\mu
   \sigma }\right)
    \right.  \nonumber\\
&&   
\left. 
   +4 \left(p_1^{\sigma } p_3^{\mu } g^{\alpha \beta
   } g^{\lambda \nu }+p_1^{\beta } p_2^{\mu } g^{\alpha \nu }
   g^{\lambda \sigma }+p_1^{\alpha } p_2^{\nu } g^{\beta \mu }
   g^{\lambda \sigma }+p_2^{\lambda } p_3^{\beta } g^{\alpha \sigma
   } g^{\mu \nu }+p_2^{\sigma } p_3^{\alpha } g^{\beta \lambda }
   g^{\mu \nu }+p_1^{\lambda } p_3^{\nu } g^{\alpha \beta } g^{\mu
   \sigma }\right)
    \right.  \nonumber\\
&&   
\left. 
   -4 \left(p_1^{\beta } p_3^{\nu } g^{\alpha \sigma
   } g^{\lambda \mu }+p_1^{\sigma } \left(p_2^{\nu } g^{\alpha \mu
   } g^{\beta \lambda }+p_3^{\beta } g^{\alpha \nu } g^{\lambda \mu
   }\right)+p_2^{\sigma } \left(p_3^{\nu } g^{\alpha \lambda }
   g^{\beta \mu }+p_1^{\beta } g^{\alpha \mu } g^{\lambda \nu
   }\right)+p_2^{\nu } p_3^{\beta } g^{\alpha \lambda } g^{\mu
   \sigma }\right)
    \right.  \nonumber\\
&&   
\left. 
   -6 \left(p_2^{\nu } p_3^{\beta } g^{\alpha \sigma
   } g^{\lambda \mu }+p_2^{\sigma } \left(p_3^{\nu } g^{\alpha \mu
   } g^{\beta \lambda }+p_1^{\beta } g^{\alpha \nu } g^{\lambda \mu
   }\right)+p_1^{\sigma } \left(p_2^{\nu } g^{\alpha \lambda }
   g^{\beta \mu }+p_3^{\beta } g^{\alpha \mu } g^{\lambda \nu
   }\right)+p_1^{\beta } p_3^{\nu } g^{\alpha \lambda } g^{\mu
   \sigma }\right)
    \right.  \nonumber\\
&&   
\left. 
   +8 \Lambda  \left(g^{\alpha \sigma } g^{\beta \mu
   } g^{\lambda \nu }+g^{\alpha \nu } g^{\beta \lambda } g^{\mu
   \sigma }\right)-2 \left(p_2^{\nu } p_3^{\mu } g^{\alpha \sigma }
   g^{\beta \lambda }+p_2^{\mu } p_3^{\nu } g^{\alpha \sigma }
   g^{\beta \lambda }+p_1^{\sigma } p_2^{\lambda } g^{\alpha \nu }
   g^{\beta \mu }+p_1^{\lambda } p_2^{\sigma } g^{\alpha \nu }
   g^{\beta \mu }
    \right. \right. \nonumber\\
&&   
\left. \left.
   +p_1^{\beta } p_3^{\alpha } g^{\lambda \nu }
   g^{\mu \sigma }+p_1^{\alpha } p_3^{\beta } g^{\lambda \nu }
   g^{\mu \sigma }\right)-2 \left(p_1^{\sigma } p_3^{\alpha }
   g^{\beta \mu } g^{\lambda \nu }+g^{\alpha \sigma }
   \left(p_1^{\lambda } p_2^{\nu } g^{\beta \mu }+p_2^{\mu }
   p_3^{\beta } g^{\lambda \nu }\right)+p_1^{\alpha } p_3^{\nu }
   g^{\beta \lambda } g^{\mu \sigma }
    \right. \right. \nonumber\\
&&   
\left. \left.
   +g^{\alpha \nu }
   \left(p_2^{\sigma } p_3^{\mu } g^{\beta \lambda }+p_1^{\beta }
   p_2^{\lambda } g^{\mu \sigma }\right)\right)+2 \left(p_2^{\nu }
   p_3^{\sigma } g^{\alpha \mu } g^{\beta \lambda }+p_1^{\sigma }
   p_2^{\beta } g^{\alpha \nu } g^{\lambda \mu }+p_2^{\beta }
   p_3^{\nu } g^{\alpha \sigma } g^{\lambda \mu }+p_1^{\beta }
   p_3^{\sigma } g^{\alpha \mu } g^{\lambda \nu }
    \right. \right. \nonumber\\
&&   
\left. \left.
   +p_1^{\nu }
   g^{\alpha \lambda } \left(p_2^{\sigma } g^{\beta \mu
   }+p_3^{\beta } g^{\mu \sigma }\right)\right)+2 \left(p_1^{\mu }
   \left(p_2^{\sigma } g^{\alpha \beta } g^{\lambda \nu
   }+p_3^{\beta } g^{\alpha \nu } g^{\lambda \sigma
   }\right)+p_2^{\alpha } \left(p_3^{\nu } g^{\beta \mu }
   g^{\lambda \sigma }+p_1^{\sigma } g^{\beta \lambda } g^{\mu \nu
   }\right)
    \right. \right. \nonumber\\
&&   
\left. \left.
   +p_3^{\lambda } \left(p_1^{\beta } g^{\alpha \sigma }
   g^{\mu \nu }+p_2^{\nu } g^{\alpha \beta } g^{\mu \sigma
   }\right)\right)-2 \left(p_1^{\mu } \left(p_3^{\sigma } g^{\alpha
   \beta } g^{\lambda \nu }+p_2^{\beta } g^{\alpha \nu } g^{\lambda
   \sigma }\right)+\left(p_2^{\beta } p_3^{\lambda } g^{\alpha
   \sigma }+p_2^{\alpha } p_3^{\sigma } g^{\beta \lambda }\right)
   g^{\mu \nu }
    \right. \right. \nonumber\\
&&   
\left. \left.
   +p_1^{\nu } \left(p_2^{\alpha } g^{\beta \mu }
   g^{\lambda \sigma }+p_3^{\lambda } g^{\alpha \beta } g^{\mu
   \sigma }\right)\right)+2 \left(p_1^{\mu } \left(p_1^{\sigma }
   g^{\alpha \beta } g^{\lambda \nu }+p_1^{\beta } g^{\alpha \nu }
   g^{\lambda \sigma }\right)+p_2^{\alpha } \left(p_2^{\nu }
   g^{\beta \mu } g^{\lambda \sigma }+p_2^{\sigma } g^{\beta
   \lambda } g^{\mu \nu }\right)
    \right. \right. \nonumber\\
&&   
\left. \left.
   +p_3^{\lambda } \left(p_3^{\beta }
   g^{\alpha \sigma } g^{\mu \nu }+p_3^{\nu } g^{\alpha \beta }
   g^{\mu \sigma }\right)\right)-4 \left(p_1^{\sigma } p_3^{\nu }
   \left(g^{\alpha \mu } g^{\beta \lambda }+g^{\alpha \lambda }
   g^{\beta \mu }\right)+p_2^{\sigma } p_3^{\beta } \left(g^{\alpha
   \nu } g^{\lambda \mu }+g^{\alpha \mu } g^{\lambda \nu
   }\right)
    \right. \right. \nonumber\\
&&   
\left. \left.
   +p_1^{\beta } p_2^{\nu } \left(g^{\alpha \sigma }
   g^{\lambda \mu }+g^{\alpha \lambda } g^{\mu \sigma
   }\right)\right)-4 \left(p_2^{\nu } p_2^{\sigma } \left(g^{\alpha
   \mu } g^{\beta \lambda }+g^{\alpha \lambda } g^{\beta \mu
   }\right)+p_1^{\beta } p_1^{\sigma } \left(g^{\alpha \nu }
   g^{\lambda \mu }+g^{\alpha \mu } g^{\lambda \nu
   }\right)
    \right. \right. \nonumber\\
&&   
\left. \left.
   +p_3^{\beta } p_3^{\nu } \left(g^{\alpha \sigma }
   g^{\lambda \mu }+g^{\alpha \lambda } g^{\mu \sigma
   }\right)\right)+2 \left(p_2^{\mu } p_3^{\sigma } g^{\alpha \nu }
   g^{\beta \lambda }+p_2^{\beta } p_3^{\mu } g^{\alpha \sigma }
   g^{\lambda \nu }+p_1^{\alpha } p_3^{\sigma } g^{\beta \mu }
   g^{\lambda \nu }+p_1^{\lambda } p_2^{\beta } g^{\alpha \nu }
   g^{\mu \sigma }
    \right. \right. \nonumber\\
&&   
\left. \left.
   +p_1^{\nu } \left(p_2^{\lambda } g^{\alpha \sigma
   } g^{\beta \mu }+p_3^{\alpha } g^{\beta \lambda } g^{\mu \sigma
   }\right)\right)+16 \left(p_3^{\alpha } p_3^{\mu } g^{\beta \nu }
   g^{\lambda \sigma }+p_1^{\alpha } p_1^{\lambda } g^{\beta \sigma
   } g^{\mu \nu }+p_2^{\lambda } p_2^{\mu } g^{\alpha \beta }
   g^{\nu \sigma }\right)-8 \left(p_1^{\alpha } p_2^{\mu } g^{\beta
   \nu } g^{\lambda \sigma }
    \right. \right. \nonumber\\
&&   
\left. \left.
   +p_2^{\lambda } p_3^{\alpha } g^{\beta
   \sigma } g^{\mu \nu }+p_1^{\lambda } p_3^{\mu } g^{\alpha \beta
   } g^{\nu \sigma }\right)+2 \left(p_1^{\alpha } p_1^{\mu }
   g^{\beta \nu } g^{\lambda \sigma }+p_3^{\alpha } p_3^{\lambda }
   g^{\beta \sigma } g^{\mu \nu }+p_2^{\alpha } \left(p_2^{\mu }
   g^{\beta \nu } g^{\lambda \sigma }+p_2^{\lambda } g^{\beta
   \sigma } g^{\mu \nu }\right)
    \right. \right. \nonumber\\
&&   
\left. \left.
   +p_1^{\lambda } p_1^{\mu } g^{\alpha
   \beta } g^{\nu \sigma }+p_3^{\lambda } p_3^{\mu } g^{\alpha
   \beta } g^{\nu \sigma }\right)+2 \left(p_3^{\alpha }
   \left(p_2^{\mu } g^{\beta \nu } g^{\lambda \sigma }+p_1^{\lambda
   } g^{\beta \sigma } g^{\mu \nu }\right)+p_1^{\alpha }
   \left(p_3^{\mu } g^{\beta \nu } g^{\lambda \sigma }+p_2^{\lambda
   } g^{\beta \sigma } g^{\mu \nu }\right)
    \right. \right. \nonumber\\
&&   
\left. \left.
   +\left(p_1^{\lambda }
   p_2^{\mu }+p_2^{\lambda } p_3^{\mu }\right) g^{\alpha \beta }
   g^{\nu \sigma }\right)-4 \left(p_2^{\mu } p_3^{\lambda }
   g^{\alpha \sigma } g^{\beta \nu }+p_1^{\alpha } p_3^{\lambda }
   g^{\mu \sigma } g^{\beta \nu }+p_1^{\mu } g^{\beta \sigma }
   \left(p_2^{\lambda } g^{\alpha \nu }+p_3^{\alpha } g^{\lambda
   \nu }\right)+p_2^{\alpha } p_3^{\mu } g^{\beta \lambda } g^{\nu
   \sigma }
    \right. \right. \nonumber\\
&&   
\left. \left.
   +p_1^{\lambda } p_2^{\alpha } g^{\beta \mu } g^{\nu
   \sigma }\right)-8 \left(p_1^{\lambda } p_1^{\mu } g^{\alpha \nu
   } g^{\beta \sigma }+p_1^{\alpha } p_1^{\mu } g^{\lambda \nu }
   g^{\beta \sigma }+p_3^{\lambda } g^{\beta \nu } \left(p_3^{\mu }
   g^{\alpha \sigma }+p_3^{\alpha } g^{\mu \sigma
   }\right)+p_2^{\alpha } p_2^{\mu } g^{\beta \lambda } g^{\nu
   \sigma }+p_2^{\alpha } p_2^{\lambda } g^{\beta \mu } g^{\nu
   \sigma }\right)
    \right.  \nonumber\\
&&   
\left. 
   +4 \left(p_3^{\lambda } p_3^{\sigma } g^{\alpha
   \mu } g^{\beta \nu }+p_1^{\mu } p_1^{\nu } g^{\alpha \lambda }
   g^{\beta \sigma }+p_2^{\alpha } p_2^{\beta } g^{\lambda \mu }
   g^{\nu \sigma }\right)-2 \left(p_1^{\lambda } p_3^{\sigma }
   g^{\alpha \mu } g^{\beta \nu }+p_2^{\lambda } p_3^{\sigma }
   g^{\alpha \mu } g^{\beta \nu }+p_1^{\nu } p_2^{\mu } g^{\alpha
   \lambda } g^{\beta \sigma }
    \right. \right. \nonumber\\
&&   
\left. \left.
   +p_1^{\nu } p_3^{\mu } g^{\alpha
   \lambda } g^{\beta \sigma }+p_1^{\alpha } p_2^{\beta }
   g^{\lambda \mu } g^{\nu \sigma }+p_2^{\beta } p_3^{\alpha }
   g^{\lambda \mu } g^{\nu \sigma }\right)+8 \left(p_1^{\sigma }
   p_2^{\lambda } g^{\alpha \mu } g^{\beta \nu }+p_1^{\lambda }
   p_2^{\sigma } g^{\alpha \mu } g^{\beta \nu }+p_2^{\nu } p_3^{\mu
   } g^{\alpha \lambda } g^{\beta \sigma }
   +p_2^{\mu } p_3^{\nu }
   g^{\alpha \lambda } g^{\beta \sigma }
   \right. \right. \nonumber\\
&&   
\left. \left.
   +p_1^{\beta } p_3^{\alpha }
   g^{\lambda \mu } g^{\nu \sigma }+p_1^{\alpha } p_3^{\beta }
   g^{\lambda \mu } g^{\nu \sigma }\right)+8 \left(p_1^{\lambda }
   p_1^{\sigma } g^{\alpha \mu } g^{\beta \nu }+p_2^{\lambda }
   p_2^{\sigma } g^{\alpha \mu } g^{\beta \nu }+p_2^{\mu } p_2^{\nu
   } g^{\alpha \lambda } g^{\beta \sigma }+p_3^{\mu } p_3^{\nu }
   g^{\alpha \lambda } g^{\beta \sigma }+p_1^{\alpha } p_1^{\beta }
   g^{\lambda \mu } g^{\nu \sigma }
   \right. \right. \nonumber\\
&&   
\left. \left.
   +p_3^{\alpha } p_3^{\beta }
   g^{\lambda \mu } g^{\nu \sigma }\right)+10
   \left(\left(p_1^{\sigma }+p_2^{\sigma }\right) p_3^{\lambda }
   g^{\alpha \mu } g^{\beta \nu }+p_1^{\mu } \left(p_2^{\nu
   }+p_3^{\nu }\right) g^{\alpha \lambda } g^{\beta \sigma
   }+p_2^{\alpha } \left(p_1^{\beta }+p_3^{\beta }\right)
   g^{\lambda \mu } g^{\nu \sigma }\right)
   \right.  \nonumber\\
&&   
\left. 
   -4 \Lambda 
   \left(g^{\alpha \mu } g^{\beta \nu } g^{\lambda \sigma
   }+g^{\alpha \lambda } g^{\beta \sigma } g^{\mu \nu }+g^{\alpha
   \beta } g^{\lambda \mu } g^{\nu \sigma }\right)+4
   \left(p_2^{\alpha } \left(p_3^{\sigma } g^{\beta \nu }+p_1^{\nu
   } g^{\beta \sigma }\right) g^{\lambda \mu }+p_1^{\mu } g^{\alpha
   \lambda } \left(p_3^{\sigma } g^{\beta \nu }+p_2^{\beta } g^{\nu
   \sigma }\right)
   \right. \right. \nonumber\\
&&   
\left. \left.
   +p_3^{\lambda } g^{\alpha \mu } \left(p_1^{\nu }
   g^{\beta \sigma }+p_2^{\beta } g^{\nu \sigma }\right)\right)-8
   \left(\left(p_3^{\alpha } p_3^{\sigma } g^{\beta \nu
   }+p_1^{\alpha } p_1^{\nu } g^{\beta \sigma }\right) g^{\lambda
   \mu }+g^{\alpha \mu } \left(p_1^{\lambda } p_1^{\nu } g^{\beta
   \sigma }+p_2^{\beta } p_2^{\lambda } g^{\nu \sigma
   }\right)
   \right. \right. \nonumber\\
&&   
\left. \left.
   +g^{\alpha \lambda } \left(p_3^{\mu } p_3^{\sigma }
   g^{\beta \nu }+p_2^{\beta } p_2^{\mu } g^{\nu \sigma
   }\right)\right)-10 \left(p_2^{\alpha } \left(p_1^{\sigma }
   g^{\beta \nu }+p_3^{\nu } g^{\beta \sigma }\right) g^{\lambda
   \mu }+p_3^{\lambda } g^{\alpha \mu } \left(p_2^{\nu } g^{\beta
   \sigma }+p_1^{\beta } g^{\nu \sigma }\right)
   \right. \right. \nonumber\\
&&   
\left. \left.
   +p_1^{\mu }
   g^{\alpha \lambda } \left(p_2^{\sigma } g^{\beta \nu
   }+p_3^{\beta } g^{\nu \sigma }\right)\right)-4
   \left(\left(p_2^{\sigma } p_3^{\alpha } g^{\beta \nu
   }+p_1^{\alpha } p_2^{\nu } g^{\beta \sigma }\right) g^{\lambda
   \mu }+g^{\alpha \lambda } \left(p_1^{\sigma } p_3^{\mu }
   g^{\beta \nu }+p_1^{\beta } p_2^{\mu } g^{\nu \sigma
   }\right)
   \right. \right. \nonumber\\
&&   
\left. \left.
   +g^{\alpha \mu } \left(p_1^{\lambda } p_3^{\nu }
   g^{\beta \sigma }+p_2^{\lambda } p_3^{\beta } g^{\nu \sigma
   }\right)\right)-4 \left(\left(p_1^{\sigma } p_3^{\alpha }
   g^{\beta \nu }+p_1^{\alpha } p_3^{\nu } g^{\beta \sigma }\right)
   g^{\lambda \mu }+g^{\alpha \mu } \left(p_1^{\lambda } p_2^{\nu }
   g^{\beta \sigma }+p_1^{\beta } p_2^{\lambda } g^{\nu \sigma
   }\right)
   \right. \right. \nonumber\\
&&   
\left. \left.
   +g^{\alpha \lambda } \left(p_2^{\sigma } p_3^{\mu }
   g^{\beta \nu }+p_2^{\mu } p_3^{\beta } g^{\nu \sigma
   }\right)\right)+10 \left(\left(p_1^{\alpha } p_2^{\sigma }
   g^{\beta \nu }+p_2^{\nu } p_3^{\alpha } g^{\beta \sigma }\right)
   g^{\lambda \mu }+g^{\alpha \mu } \left(p_2^{\lambda } p_3^{\nu }
   g^{\beta \sigma }+p_1^{\lambda } p_3^{\beta } g^{\nu \sigma
   }\right)
   \right. \right. \nonumber\\
&&   
\left. \left.
   +g^{\alpha \lambda } \left(p_1^{\sigma } p_2^{\mu }
   g^{\beta \nu }+p_1^{\beta } p_3^{\mu } g^{\nu \sigma
   }\right)\right)+8 \left(p_2^{\alpha } \left(p_3^{\mu } g^{\beta
   \nu } g^{\lambda \sigma }+p_1^{\lambda } g^{\beta \sigma }
   g^{\mu \nu }\right)+p_1^{\mu } \left(p_3^{\alpha } g^{\beta \nu
   } g^{\lambda \sigma }+p_2^{\lambda } g^{\alpha \beta } g^{\nu
   \sigma }\right)
   \right. \right. \nonumber\\
&&   
\left. \left.
   +p_3^{\lambda } \left(p_1^{\alpha } g^{\beta
   \sigma } g^{\mu \nu }+p_2^{\mu } g^{\alpha \beta } g^{\nu \sigma
   }\right)\right)+12 \left(p_2^{\alpha } p_3^{\lambda } g^{\beta
   \sigma } g^{\mu \nu }+p_1^{\mu } \left(p_2^{\alpha } g^{\beta
   \nu } g^{\lambda \sigma }+p_3^{\lambda } g^{\alpha \beta }
   g^{\nu \sigma }\right)\right)-6 \left(p_2^{\mu } p_3^{\lambda }
   g^{\alpha \nu } g^{\beta \sigma }
   \right. \right. \nonumber\\
&&   
\left. \left.
   +p_2^{\alpha } p_3^{\mu }
   g^{\lambda \nu } g^{\beta \sigma }+p_1^{\lambda } p_2^{\alpha }
   g^{\beta \nu } g^{\mu \sigma }+p_1^{\alpha } p_3^{\lambda }
   g^{\beta \mu } g^{\nu \sigma }+p_1^{\mu } \left(p_2^{\lambda }
   g^{\alpha \sigma } g^{\beta \nu }+p_3^{\alpha } g^{\beta \lambda
   } g^{\nu \sigma }\right)\right)-4 \left(p_2^{\lambda } p_2^{\mu
   } \left(g^{\alpha \sigma } g^{\beta \nu }+g^{\alpha \nu }
   g^{\beta \sigma }\right)
   \right. \right. \nonumber\\
&&   
\left. \left.
   +p_3^{\alpha } p_3^{\mu } \left(g^{\beta
   \sigma } g^{\lambda \nu }+g^{\beta \lambda } g^{\nu \sigma
   }\right)+p_1^{\alpha } p_1^{\lambda } \left(g^{\beta \nu }
   g^{\mu \sigma }+g^{\beta \mu } g^{\nu \sigma }\right)\right)+4
   \left(p_1^{\lambda } p_3^{\mu } \left(g^{\alpha \sigma }
   g^{\beta \nu }+g^{\alpha \nu } g^{\beta \sigma
   }\right)
   \right. \right. \nonumber\\
&&   
\left. \left.
   +p_1^{\alpha } p_2^{\mu } \left(g^{\beta \sigma }
   g^{\lambda \nu }+g^{\beta \lambda } g^{\nu \sigma
   }\right)+p_2^{\lambda } p_3^{\alpha } \left(g^{\beta \nu }
   g^{\mu \sigma }+g^{\beta \mu } g^{\nu \sigma }\right)\right)-4
   \left(p_1^{\alpha } p_3^{\mu } g^{\beta \sigma } g^{\lambda \nu
   }+p_1^{\lambda } \left(p_2^{\mu } g^{\alpha \nu } g^{\beta
   \sigma }+p_3^{\alpha } g^{\beta \nu } g^{\mu \sigma
   }\right)
   \right. \right. \nonumber\\
&&   
\left. \left.
   +p_2^{\mu } p_3^{\alpha } g^{\beta \lambda } g^{\nu
   \sigma }+p_2^{\lambda } \left(p_3^{\mu } g^{\alpha \sigma }
   g^{\beta \nu }+p_1^{\alpha } g^{\beta \mu } g^{\nu \sigma
   }\right)\right)+2 \left(p_2^{\mu } p_3^{\alpha } g^{\beta \sigma
   } g^{\lambda \nu }+p_2^{\lambda } \left(p_3^{\mu } g^{\alpha \nu
   } g^{\beta \sigma }+p_1^{\alpha } g^{\beta \nu } g^{\mu \sigma
   }\right)+p_1^{\alpha } p_3^{\mu } g^{\beta \lambda } g^{\nu
   \sigma }
   \right. \right. \nonumber\\
&&   
\left. \left.
   +p_1^{\lambda } \left(p_2^{\mu } g^{\alpha \sigma }
   g^{\beta \nu }+p_3^{\alpha } g^{\beta \mu } g^{\nu \sigma
   }\right)\right)-12 \left(p_2^{\alpha } p_3^{\lambda }
   \left(g^{\beta \nu } g^{\mu \sigma }+g^{\beta \mu } g^{\nu
   \sigma }\right)+p_1^{\mu } \left(p_3^{\lambda } \left(g^{\alpha
   \sigma } g^{\beta \nu }+g^{\alpha \nu } g^{\beta \sigma
   }\right)
   \right. \right. \right. \nonumber\\
&&   
\left. \left. \left.
   +p_2^{\alpha } \left(g^{\beta \sigma } g^{\lambda \nu
   }+g^{\beta \lambda } g^{\nu \sigma }\right)\right)\right)+4
   \left(g^{\alpha \sigma } g^{\beta \nu } g^{\lambda \mu }
   p_1\cdot p_2+g^{\alpha \mu } g^{\beta \lambda } g^{\nu \sigma }
   p_1\cdot p_3+g^{\alpha \lambda } \left(g^{\beta \nu } g^{\mu
   \sigma } p_1\cdot p_2+g^{\beta \mu } g^{\nu \sigma } p_1\cdot
   p_3\right)
   \right. \right. \nonumber\\
&&   
\left. \left.
   +g^{\alpha \nu } g^{\beta \sigma } g^{\lambda \mu }
   p_2\cdot p_3+g^{\alpha \mu } g^{\beta \sigma } g^{\lambda \nu }
   p_2\cdot p_3\right)+2 \left(g^{\alpha \nu } g^{\beta \mu }
   g^{\lambda \sigma } p_1\cdot p_2+g^{\alpha \beta } g^{\lambda
   \nu } g^{\mu \sigma } p_1\cdot p_3+g^{\alpha \sigma } g^{\beta
   \lambda } g^{\mu \nu } p_2\cdot p_3\right)\right.  \nonumber\\
&&   
\left. 
-8 \left(g^{\alpha \mu
   } g^{\beta \nu } g^{\lambda \sigma } p_1\cdot p_2+g^{\alpha
   \beta } g^{\lambda \mu } g^{\nu \sigma } p_1\cdot p_3+g^{\alpha
   \lambda } g^{\beta \sigma } g^{\mu \nu } p_2\cdot p_3\right)+8
   \left(g^{\alpha \mu } \left(g^{\beta \sigma } g^{\lambda \nu
   }
   \right. \right. \right. \nonumber\\
&&   
\left. \left. \left.
   +g^{\beta \lambda } g^{\nu \sigma }\right) p_1\cdot
   p_2+g^{\alpha \sigma } g^{\beta \nu } g^{\lambda \mu } p_1\cdot
   p_3+g^{\alpha \nu } g^{\beta \sigma } g^{\lambda \mu } p_1\cdot
   p_3+g^{\alpha \lambda } g^{\beta \nu } g^{\mu \sigma } p_2\cdot
   p_3+g^{\alpha \lambda } g^{\beta \mu } g^{\nu \sigma } p_2\cdot
   p_3\right)
   \right.  \nonumber\\
&&   
\left. 
   -4 \left(g^{\alpha \lambda } g^{\beta \sigma } g^{\mu
   \nu } \left(p_1\cdot p_2+p_1\cdot p_3\right)+g^{\alpha \beta }
   g^{\lambda \mu } g^{\nu \sigma } \left(p_1\cdot p_2+p_2\cdot
   p_3\right)+g^{\alpha \mu } g^{\beta \nu } g^{\lambda \sigma }
   \left(p_1\cdot p_3+p_2\cdot p_3\right)\right)
   \right.  \nonumber\\
&&   
\left. 
   -4 \left(g^{\alpha
   \lambda } g^{\beta \sigma } g^{\mu \nu } p_1^2+g^{\alpha \beta }
   g^{\lambda \mu } g^{\nu \sigma } p_2^2+g^{\alpha \mu } g^{\beta
   \nu } g^{\lambda \sigma } p_3^2\right)+8 \left(g^{\alpha \nu }
   g^{\beta \sigma } g^{\lambda \mu } p_1^2+g^{\alpha \lambda }
   g^{\beta \mu } g^{\nu \sigma } p_2^2+g^{\alpha \sigma } g^{\beta
   \nu } g^{\lambda \mu } p_3^2
   \right. \right. \nonumber\\
&&   
\left. \left.
   +g^{\alpha \lambda } g^{\beta \nu }
   g^{\mu \sigma } p_3^2+g^{\alpha \mu } \left(g^{\beta \sigma }
   g^{\lambda \nu } p_1^2+g^{\beta \lambda } g^{\nu \sigma }
   p_2^2\right)\right)-4 \left(g^{\alpha \mu } g^{\beta \nu }
   g^{\lambda \sigma } \left(p_1^2+p_2^2\right)
   \right. \right. \nonumber\\
&&   
\left. \left.+g^{\alpha \beta }
   g^{\lambda \mu } g^{\nu \sigma }
   \left(p_1^2+p_3^2\right)+g^{\alpha \lambda } g^{\beta \sigma }
   g^{\mu \nu } \left(p_2^2+p_3^2\right)\right)\right)
;
\label{hhh}
\end{eqnarray}

\item
Energy-momentum tensor with indices $(\mu,\nu)$ - scalars with momenta $p_1$ and $p_2$:
\begin{equation}
g^{\mu \nu } \left(m^2+p_1\cdot p_2\right)-p_2^{\mu }
   p_1^{\nu } -p_1^{\mu } p_2^{\nu } ;
\label{TSS}
\end{equation}

\item
  Energy-momentum tensor with indices $(\mu,\nu)$ - graviton with indices $(\alpha,\beta)$-
  scalars with momenta $p_1$ and $p_2$:
\begin{eqnarray}
&& \frac{1}{2} \kappa  \left(m^2 \left(-g^{\alpha \nu }\right)
   g^{\beta \mu }-m^2 g^{\alpha \mu } g^{\beta \nu }+2 m^2
   g^{\alpha \beta } g^{\mu \nu }+p_1^{\beta } p_2^{\nu } g^{\alpha
   \mu }+p_1^{\beta } p_2^{\mu } g^{\alpha \nu }+p_1^{\alpha }
   p_2^{\nu } g^{\beta \mu } \right.  \nonumber\\
   && \left. +p_1^{\nu } \left(-2 p_2^{\mu }
   g^{\alpha \beta }+p_2^{\beta } g^{\alpha \mu }+p_2^{\alpha }
   g^{\beta \mu }\right)+p_1^{\alpha } p_2^{\mu } g^{\beta \nu
   }+p_1^{\mu } \left(-2 p_2^{\nu } g^{\alpha \beta }+p_2^{\beta }
   g^{\alpha \nu }+p_2^{\alpha } g^{\beta \nu }\right)-p_2^{\alpha
   } p_1^{\beta } g^{\mu \nu }
   \right.  \nonumber\\
   && \left.
   -p_1^{\alpha } p_2^{\beta } g^{\mu
   \nu }-p_1\cdot p_2 g^{\alpha \nu } g^{\beta \mu }-p_1\cdot p_2
   g^{\alpha \mu } g^{\beta \nu }+2 p_1\cdot p_2 g^{\alpha \beta }
   g^{\mu \nu }\right);
\label{ThSS}
\end{eqnarray}

\item
  Energy-momentum tensor with indices $(\mu,\nu)$ - vector bosons with (Lorentz index, momentum)
  combinations $(\lambda,p_1)$  and $(\sigma, p_2)$:
\begin{eqnarray}
&& - M^2 g^{\lambda \sigma } g^{\mu \nu }+M^2 g^{\lambda
   \nu } g^{\mu \sigma }+M^2 g^{\lambda \mu } g^{\nu \sigma
   }+p_1^{\mu } p_2^{\nu } g^{\lambda \sigma }-p_1^{\sigma }
   \left(p_2^{\nu } g^{\lambda \mu }+p_2^{\mu } g^{\lambda \nu
   }-p_2^{\lambda } g^{\mu \nu }\right)
    \nonumber\\
 && 
   +p_1^{\nu } \left(p_2^{\mu }
   g^{\lambda \sigma }-p_2^{\lambda } g^{\mu \sigma
   }\right)  -p_2^{\lambda } p_1^{\mu } g^{\nu \sigma }-p_1\cdot p_2
   g^{\lambda \sigma } g^{\mu \nu }+p_1\cdot p_2 g^{\lambda \nu }
   g^{\mu \sigma }+p_1\cdot p_2 g^{\lambda \mu } g^{\nu \sigma };
\label{TVV}
\end{eqnarray}

\item
  Energy-momentum tensor with indices $(\mu,\nu)$ - graviton with indices $(\alpha,\beta)$
  - vector bosons with (Lorentz index, momentum) combinations $(\lambda,p_1)$  and $(\sigma, p_2)$:
\begin{eqnarray}
&& -\frac{1}{2} \kappa  \left(g^{\alpha \sigma } g^{\beta \nu }
   g^{\lambda \mu } M^2+g^{\alpha \nu } g^{\beta \sigma }
   g^{\lambda \mu } M^2+g^{\alpha \sigma } g^{\beta \mu }
   g^{\lambda \nu } M^2+g^{\alpha \mu } g^{\beta \sigma }
   g^{\lambda \nu } M^2-g^{\alpha \nu } g^{\beta \mu } g^{\lambda
   \sigma } M^2-g^{\alpha \mu } g^{\beta \nu } g^{\lambda \sigma }
   M^2 
   \right.  \nonumber\\
   && \left.
   -g^{\alpha \sigma } g^{\beta \lambda } g^{\mu \nu }
   M^2-g^{\alpha \lambda } g^{\beta \sigma } g^{\mu \nu } M^2+2
   g^{\alpha \beta } g^{\lambda \sigma } g^{\mu \nu } M^2+g^{\alpha
   \nu } g^{\beta \lambda } g^{\mu \sigma } M^2+g^{\alpha \lambda }
   g^{\beta \nu } g^{\mu \sigma } M^2-2 g^{\alpha \beta }
   g^{\lambda \nu } g^{\mu \sigma } M^2
   \right.  \nonumber\\
   && \left.
   +g^{\alpha \mu } g^{\beta
   \lambda } g^{\nu \sigma } M^2+g^{\alpha \lambda } g^{\beta \mu }
   g^{\nu \sigma } M^2-2 g^{\alpha \beta } g^{\lambda \mu } g^{\nu
   \sigma } M^2+p_1^{\mu } p_2^{\nu } g^{\alpha \sigma } g^{\beta
   \lambda }-p_1^{\mu } p_2^{\lambda } g^{\alpha \sigma } g^{\beta
   \nu }+p_1^{\mu } p_2^{\nu } g^{\alpha \lambda } g^{\beta \sigma
   } 
   \right.  \nonumber\\
   && \left.
   -p_1^{\mu } p_2^{\lambda } g^{\alpha \nu } g^{\beta \sigma
   }-p_1^{\beta } p_2^{\nu } g^{\alpha \sigma } g^{\lambda \mu
   }-p_1^{\alpha } p_2^{\nu } g^{\beta \sigma } g^{\lambda \mu
   }-p_1^{\beta } p_2^{\mu } g^{\alpha \sigma } g^{\lambda \nu
   }-p_1^{\alpha } p_2^{\mu } g^{\beta \sigma } g^{\lambda \nu }-2
   p_1^{\mu } p_2^{\nu } g^{\alpha \beta } g^{\lambda \sigma
   }+p_1^{\beta } p_2^{\nu } g^{\alpha \mu } g^{\lambda \sigma
   }
   \right.  \nonumber\\
   && \left.
   +p_1^{\mu } p_2^{\beta } g^{\alpha \nu } g^{\lambda \sigma
   }+p_1^{\beta } p_2^{\mu } g^{\alpha \nu } g^{\lambda \sigma
   }+p_1^{\alpha } p_2^{\nu } g^{\beta \mu } g^{\lambda \sigma
   }+p_1^{\mu } p_2^{\alpha } g^{\beta \nu } g^{\lambda \sigma
   }+p_1^{\alpha } p_2^{\mu } g^{\beta \nu } g^{\lambda \sigma
   }+p_1^{\beta } p_2^{\lambda } g^{\alpha \sigma } g^{\mu \nu
   }+p_1^{\alpha } p_2^{\lambda } g^{\beta \sigma } g^{\mu \nu
   }
   \right.  \nonumber\\
   && \left.
   -p_1^{\beta } p_2^{\alpha } g^{\lambda \sigma } g^{\mu \nu
   }-p_1^{\alpha } p_2^{\beta } g^{\lambda \sigma } g^{\mu \nu
   }-p_1^{\sigma } \left(-p_2^{\lambda } g^{\alpha \nu } g^{\beta
   \mu }+p_2^{\alpha } g^{\lambda \nu } g^{\beta \mu }-p_2^{\lambda
   } g^{\alpha \mu } g^{\beta \nu }+p_2^{\beta } g^{\alpha \nu }
   g^{\lambda \mu }+p_2^{\alpha } g^{\beta \nu } g^{\lambda \mu
   }
   \right. \right. \nonumber\\
   && \left.\left.
   +p_2^{\nu } \left(g^{\alpha \mu } g^{\beta \lambda }+g^{\alpha
   \lambda } g^{\beta \mu }-2 g^{\alpha \beta } g^{\lambda \mu
   }\right)+p_2^{\beta } g^{\alpha \mu } g^{\lambda \nu }+p_2^{\mu
   } \left(g^{\alpha \nu } g^{\beta \lambda }+g^{\alpha \lambda }
   g^{\beta \nu }-2 g^{\alpha \beta } g^{\lambda \nu }\right)+2
   p_2^{\lambda } g^{\alpha \beta } g^{\mu \nu }
  \right. \right. \nonumber\\
   && \left.\left.
   -p_2^{\beta }
   g^{\alpha \lambda } g^{\mu \nu }-p_2^{\alpha } g^{\beta \lambda
   } g^{\mu \nu }\right)-p_1^{\beta } p_2^{\lambda } g^{\alpha \nu
   } g^{\mu \sigma }-p_1^{\alpha } p_2^{\lambda } g^{\beta \nu }
   g^{\mu \sigma }+p_1^{\beta } p_2^{\alpha } g^{\lambda \nu }
   g^{\mu \sigma }+p_1^{\alpha } p_2^{\beta } g^{\lambda \nu }
   g^{\mu \sigma }
  \right.  \nonumber\\
   && \left.  
   +p_1^{\nu } \left(p_2^{\beta } g^{\alpha \mu }
   g^{\lambda \sigma }+p_2^{\alpha } g^{\beta \mu } g^{\lambda
   \sigma }+p_2^{\mu } \left(g^{\alpha \sigma } g^{\beta \lambda
   }+g^{\alpha \lambda } g^{\beta \sigma }-2 g^{\alpha \beta }
   g^{\lambda \sigma }\right)-p_2^{\beta } g^{\alpha \lambda }
   g^{\mu \sigma }-p_2^{\alpha } g^{\beta \lambda } g^{\mu \sigma
   }
   \right. \right. \nonumber\\
   && \left.\left.
   +p_2^{\lambda } \left(-g^{\alpha \sigma } g^{\beta \mu
   }-g^{\alpha \mu } g^{\beta \sigma }+2 g^{\alpha \beta } g^{\mu
   \sigma }\right)\right)+2 p_1^{\mu } p_2^{\lambda } g^{\alpha
   \beta } g^{\nu \sigma }-p_1^{\mu } p_2^{\beta } g^{\alpha
   \lambda } g^{\nu \sigma }-p_1^{\beta } p_2^{\lambda } g^{\alpha
   \mu } g^{\nu \sigma }-p_1^{\mu } p_2^{\alpha } g^{\beta \lambda
   } g^{\nu \sigma }
     \right.  \nonumber\\
   && \left.
   -p_1^{\alpha } p_2^{\lambda } g^{\beta \mu }
   g^{\nu \sigma }+p_1^{\beta } p_2^{\alpha } g^{\lambda \mu }
   g^{\nu \sigma }+p_1^{\alpha } p_2^{\beta } g^{\lambda \mu }
   g^{\nu \sigma }+g^{\alpha \sigma } g^{\beta \nu } g^{\lambda \mu
   } p_1\cdot p_2+g^{\alpha \nu } g^{\beta \sigma } g^{\lambda \mu
   } p_1\cdot p_2+g^{\alpha \sigma } g^{\beta \mu } g^{\lambda \nu
   } p_1\cdot p_2
    \right.  \nonumber\\
   && \left.
   +g^{\alpha \mu } g^{\beta \sigma } g^{\lambda \nu
   } p_1\cdot p_2-g^{\alpha \nu } g^{\beta \mu } g^{\lambda \sigma
   } p_1\cdot p_2-g^{\alpha \mu } g^{\beta \nu } g^{\lambda \sigma
   } p_1\cdot p_2-g^{\alpha \sigma } g^{\beta \lambda } g^{\mu \nu
   } p_1\cdot p_2-g^{\alpha \lambda } g^{\beta \sigma } g^{\mu \nu
   } p_1\cdot p_2
    \right.  \nonumber\\
   && \left.
   +2 g^{\alpha \beta } g^{\lambda \sigma } g^{\mu
   \nu } p_1\cdot p_2+g^{\alpha \nu } g^{\beta \lambda } g^{\mu
   \sigma } p_1\cdot p_2+g^{\alpha \lambda } g^{\beta \nu } g^{\mu
   \sigma } p_1\cdot p_2-2 g^{\alpha \beta } g^{\lambda \nu }
   g^{\mu \sigma } p_1\cdot p_2
    \right.  \nonumber\\
   && \left.
   +g^{\alpha \mu } g^{\beta \lambda }
   g^{\nu \sigma } p_1\cdot p_2+g^{\alpha \lambda } g^{\beta \mu }
   g^{\nu \sigma } p_1\cdot p_2-2 g^{\alpha \beta } g^{\lambda \mu
   } g^{\nu \sigma } p_1\cdot p_2\right) ;
\label{ThVV}
\end{eqnarray}

\item
  Energy-momentum tensor with indices $(\mu,\nu)$ - gravitons with (Lorentz indices, momentum)
  combinations $(\lambda,\sigma, p_1)$  and $(\alpha,\beta, p_2)$:
\begin{eqnarray}
&&
\frac{1}{8} \left(-4 p_1^{\sigma } p_2^{\lambda } g^{\alpha \nu }
   g^{\beta \mu }-4 p_1^{\lambda } p_2^{\sigma } g^{\alpha \nu }
   g^{\beta \mu }+4 p_1^{\sigma } p_2^{\alpha } g^{\lambda \nu }
   g^{\beta \mu }-2 p_1^{\alpha } p_2^{\nu } g^{\lambda \sigma }
   g^{\beta \mu }-2 p_1^{\alpha } p_2^{\nu } g^{\sigma \lambda }
   g^{\beta \mu }+4 p_1^{\lambda } p_2^{\alpha } g^{\sigma \nu }
   g^{\beta \mu }  
   \right.  \nonumber\\
   && \left.
   -4 g^{\alpha \sigma } g^{\lambda \nu } p_1\cdot
   p_2 g^{\beta \mu }+6 g^{\alpha \nu } g^{\lambda \sigma }
   p_1\cdot p_2 g^{\beta \mu }+6 g^{\alpha \nu } g^{\sigma \lambda
   } p_1\cdot p_2 g^{\beta \mu }-4 g^{\alpha \lambda } g^{\sigma
   \nu } p_1\cdot p_2 g^{\beta \mu }-4 p_1^{\sigma } p_2^{\lambda }
   g^{\alpha \mu } g^{\beta \nu }
   \right.  \nonumber\\
   && \left.
   -4 p_1^{\lambda } p_2^{\sigma }
   g^{\alpha \mu } g^{\beta \nu }-4 p_1^{\sigma } p_2^{\nu }
   g^{\alpha \beta } g^{\lambda \mu }+4 p_1^{\sigma } p_2^{\beta }
   g^{\alpha \nu } g^{\lambda \mu }+4 p_1^{\beta } p_2^{\nu }
   g^{\alpha \sigma } g^{\lambda \mu }-4 p_1^{\sigma } p_2^{\nu }
   g^{\beta \alpha } g^{\lambda \mu }+4 p_1^{\sigma } p_2^{\alpha }
   g^{\beta \nu } g^{\lambda \mu }
   \right.  \nonumber\\
   && \left.
   +4 p_1^{\alpha } p_2^{\nu }
   g^{\beta \sigma } g^{\lambda \mu }-4 p_1^{\sigma } p_2^{\mu }
   g^{\alpha \beta } g^{\lambda \nu }+4 p_1^{\sigma } p_2^{\beta }
   g^{\alpha \mu } g^{\lambda \nu }+4 p_1^{\beta } p_2^{\mu }
   g^{\alpha \sigma } g^{\lambda \nu }-4 p_1^{\sigma } p_2^{\mu }
   g^{\beta \alpha } g^{\lambda \nu }+4 p_1^{\alpha } p_2^{\mu }
   g^{\beta \sigma } g^{\lambda \nu } 
   \right.  \nonumber\\
   && \left.
   -2 p_1^{\beta } p_2^{\nu }
   g^{\alpha \mu } g^{\lambda \sigma }-2 p_1^{\beta } p_2^{\mu }
   g^{\alpha \nu } g^{\lambda \sigma }-2 p_1^{\alpha } p_2^{\mu }
   g^{\beta \nu } g^{\lambda \sigma }+2 p_1^{\sigma } p_2^{\lambda
   } g^{\alpha \beta } g^{\mu \nu }+2 p_1^{\lambda } p_2^{\sigma }
   g^{\alpha \beta } g^{\mu \nu }-2 p_1^{\beta } p_2^{\sigma }
   g^{\alpha \lambda } g^{\mu \nu }
   \right.  \nonumber\\
   && \left.
   -2 p_1^{\beta } p_2^{\lambda }
   g^{\alpha \sigma } g^{\mu \nu }+2 p_1^{\sigma } p_2^{\lambda }
   g^{\beta \alpha } g^{\mu \nu }+2 p_1^{\lambda } p_2^{\sigma }
   g^{\beta \alpha } g^{\mu \nu }-2 p_1^{\alpha } p_2^{\sigma }
   g^{\beta \lambda } g^{\mu \nu }-2 p_1^{\alpha } p_2^{\lambda }
   g^{\beta \sigma } g^{\mu \nu }+2 p_1^{\beta } p_2^{\alpha }
   g^{\lambda \sigma } g^{\mu \nu }
   \right.  \nonumber\\
   && \left.
   +2 p_1^{\alpha } p_2^{\beta }
   g^{\lambda \sigma } g^{\mu \nu }-2 p_1^{\beta } p_2^{\alpha }
   g^{\lambda \nu } g^{\mu \sigma }-2 p_1^{\alpha } p_2^{\beta }
   g^{\lambda \nu } g^{\mu \sigma }+2 p_1^{\sigma } p_2^{\lambda }
   g^{\alpha \beta } g^{\nu \mu }+2 p_1^{\lambda } p_2^{\sigma }
   g^{\alpha \beta } g^{\nu \mu }-2 p_1^{\beta } p_2^{\sigma }
   g^{\alpha \lambda } g^{\nu \mu }
   \right.  \nonumber\\
   && \left.
   -2 p_1^{\beta } p_2^{\lambda }
   g^{\alpha \sigma } g^{\nu \mu }+2 p_1^{\sigma } p_2^{\lambda }
   g^{\beta \alpha } g^{\nu \mu }+2 p_1^{\lambda } p_2^{\sigma }
   g^{\beta \alpha } g^{\nu \mu }-2 p_1^{\alpha } p_2^{\sigma }
   g^{\beta \lambda } g^{\nu \mu }-2 p_1^{\alpha } p_2^{\lambda }
   g^{\beta \sigma } g^{\nu \mu }+2 p_1^{\beta } p_2^{\alpha }
   g^{\lambda \sigma } g^{\nu \mu }
   \right.  \nonumber\\
   && \left.
   +2 p_1^{\alpha } p_2^{\beta }
   g^{\lambda \sigma } g^{\nu \mu }-2 p_1^{\beta } p_2^{\alpha }
   g^{\lambda \mu } g^{\nu \sigma }-2 p_1^{\alpha } p_2^{\beta }
   g^{\lambda \mu } g^{\nu \sigma }-2 p_1^{\beta } p_2^{\nu }
   g^{\alpha \mu } g^{\sigma \lambda }-2 p_1^{\beta } p_2^{\mu }
   g^{\alpha \nu } g^{\sigma \lambda }-2 p_1^{\alpha } p_2^{\mu }
   g^{\beta \nu } g^{\sigma \lambda }
   \right.  \nonumber\\
   && \left.
   +2 p_1^{\beta } p_2^{\alpha }
   g^{\mu \nu } g^{\sigma \lambda }+2 p_1^{\alpha } p_2^{\beta }
   g^{\mu \nu } g^{\sigma \lambda }+2 p_1^{\beta } p_2^{\alpha }
   g^{\nu \mu } g^{\sigma \lambda }+2 p_1^{\alpha } p_2^{\beta }
   g^{\nu \mu } g^{\sigma \lambda }-4 p_1^{\lambda } p_2^{\nu }
   g^{\alpha \beta } g^{\sigma \mu }+4 p_1^{\beta } p_2^{\nu }
   g^{\alpha \lambda } g^{\sigma \mu }
   \right.  \nonumber\\
   && \left.
   +4 p_1^{\lambda } p_2^{\beta
   } g^{\alpha \nu } g^{\sigma \mu }-4 p_1^{\lambda } p_2^{\nu }
   g^{\beta \alpha } g^{\sigma \mu }+4 p_1^{\alpha } p_2^{\nu }
   g^{\beta \lambda } g^{\sigma \mu }+4 p_1^{\lambda } p_2^{\alpha
   } g^{\beta \nu } g^{\sigma \mu }-2 p_1^{\beta } p_2^{\alpha }
   g^{\nu \lambda } g^{\sigma \mu }-2 p_1^{\alpha } p_2^{\beta }
   g^{\nu \lambda } g^{\sigma \mu }\right.  \nonumber\\
   && \left.
   -2 p_1^{\nu } \left(-2
   p_2^{\lambda } g^{\alpha \sigma } g^{\beta \mu }+2 p_2^{\alpha }
   g^{\lambda \sigma } g^{\beta \mu }+2 p_2^{\alpha } g^{\sigma
   \lambda } g^{\beta \mu }-2 p_2^{\lambda } g^{\alpha \mu }
   g^{\beta \sigma }
   \right. \right.  \nonumber\\
   && \left. \left. 
   +p_2^{\sigma } \left(-2 g^{\alpha \mu }
   g^{\beta \lambda }-2 g^{\alpha \lambda } g^{\beta \mu
   }+\left(g^{\alpha \beta }+g^{\beta \alpha }\right) g^{\lambda
   \mu }\right)+2 p_2^{\beta } g^{\alpha \mu } g^{\lambda \sigma
   }+2 p_2^{\beta } g^{\alpha \mu } g^{\sigma \lambda }
   \right. \right.  \nonumber\\
   && \left. \left.
   +2 p_2^{\mu
   } \left(g^{\alpha \sigma } g^{\beta \lambda }+g^{\alpha \lambda
   } g^{\beta \sigma }-\left(g^{\alpha \beta }+g^{\beta \alpha
   }\right) \left(g^{\lambda \sigma }+g^{\sigma \lambda
   }\right)\right)+p_2^{\lambda } g^{\alpha \beta } g^{\sigma \mu
   }+p_2^{\lambda } g^{\beta \alpha } g^{\sigma \mu }\right)-4
   p_1^{\lambda } p_2^{\mu } g^{\alpha \beta } g^{\sigma \nu }
   \right.  \nonumber\\
   && \left.
   +4
   p_1^{\beta } p_2^{\mu } g^{\alpha \lambda } g^{\sigma \nu }+4
   p_1^{\lambda } p_2^{\beta } g^{\alpha \mu } g^{\sigma \nu }-4
   p_1^{\lambda } p_2^{\mu } g^{\beta \alpha } g^{\sigma \nu }+4
   p_1^{\alpha } p_2^{\mu } g^{\beta \lambda } g^{\sigma \nu }-2
   p_1^{\beta } p_2^{\alpha } g^{\mu \lambda } g^{\sigma \nu }-2
   p_1^{\alpha } p_2^{\beta } g^{\mu \lambda } g^{\sigma \nu }
   \right.  \nonumber\\
   && \left.
   -2 p_1^{\mu } \left(-2 p_2^{\lambda } g^{\alpha \sigma } g^{\beta
   \nu }+2 p_2^{\alpha } g^{\lambda \sigma } g^{\beta \nu }+2
   p_2^{\alpha } g^{\sigma \lambda } g^{\beta \nu }-2 p_2^{\lambda
   } g^{\alpha \nu } g^{\beta \sigma }
   \right. \right. \nonumber\\
   && \left.\left.
   +p_2^{\sigma } \left(-2
   g^{\alpha \nu } g^{\beta \lambda }-2 g^{\alpha \lambda }
   g^{\beta \nu }+\left(g^{\alpha \beta }+g^{\beta \alpha }\right)
   g^{\lambda \nu }\right)+2 p_2^{\beta } g^{\alpha \nu }
   g^{\lambda \sigma }+2 p_2^{\beta } g^{\alpha \nu } g^{\sigma
   \lambda }
   \right. \right. \nonumber\\
   && \left. \left.
   +2 p_2^{\nu } \left(g^{\alpha \sigma } g^{\beta \lambda
   }+g^{\alpha \lambda } g^{\beta \sigma }-\left(g^{\alpha \beta
   }+g^{\beta \alpha }\right) \left(g^{\lambda \sigma }+g^{\sigma
   \lambda }\right)\right)+p_2^{\lambda } g^{\alpha \beta }
   g^{\sigma \nu }+p_2^{\lambda } g^{\beta \alpha } g^{\sigma \nu
   }\right)-4 g^{\alpha \sigma } g^{\beta \nu } g^{\lambda \mu }
   p_1\cdot p_2  
   \right.  \nonumber\\
   && \left.
   -4 g^{\alpha \nu } g^{\beta \sigma } g^{\lambda \mu
   } p_1\cdot p_2
   -4 g^{\alpha \mu } g^{\beta \sigma } g^{\lambda
   \nu } p_1\cdot p_2+6 g^{\alpha \mu } g^{\beta \nu } g^{\lambda
   \sigma } p_1\cdot p_2+2 g^{\alpha \sigma } g^{\beta \lambda }
   g^{\mu \nu } p_1\cdot p_2+2 g^{\alpha \lambda } g^{\beta \sigma
   } g^{\mu \nu } p_1\cdot p_2 
   \right.  \nonumber\\
   && \left.
   -3 g^{\alpha \beta } g^{\lambda
   \sigma } g^{\mu \nu } p_1\cdot p_2-3 g^{\beta \alpha }
   g^{\lambda \sigma } g^{\mu \nu } p_1\cdot p_2+3 g^{\alpha \beta
   } g^{\lambda \nu } g^{\mu \sigma } p_1\cdot p_2+3 g^{\beta
   \alpha } g^{\lambda \nu } g^{\mu \sigma } p_1\cdot p_2+2
   g^{\alpha \sigma } g^{\beta \lambda } g^{\nu \mu } p_1\cdot
   p_2
   \right.  \nonumber\\
   && \left.
   +2 g^{\alpha \lambda } g^{\beta \sigma } g^{\nu \mu }
   p_1\cdot p_2-3 g^{\alpha \beta } g^{\lambda \sigma } g^{\nu \mu
   } p_1\cdot p_2-3 g^{\beta \alpha } g^{\lambda \sigma } g^{\nu
   \mu } p_1\cdot p_2+3 g^{\alpha \beta } g^{\lambda \mu } g^{\nu
   \sigma } p_1\cdot p_2+3 g^{\beta \alpha } g^{\lambda \mu }
   g^{\nu \sigma } p_1\cdot p_2
   \right.  \nonumber\\
   && \left.
   +6 g^{\alpha \mu } g^{\beta \nu }
   g^{\sigma \lambda } p_1\cdot p_2-3 g^{\alpha \beta } g^{\mu \nu
   } g^{\sigma \lambda } p_1\cdot p_2-3 g^{\beta \alpha } g^{\mu
   \nu } g^{\sigma \lambda } p_1\cdot p_2-3 g^{\alpha \beta }
   g^{\nu \mu } g^{\sigma \lambda } p_1\cdot p_2-3 g^{\beta \alpha
   } g^{\nu \mu } g^{\sigma \lambda } p_1\cdot p_2
   \right.  \nonumber\\
   && \left.
   -4 g^{\alpha \nu
   } g^{\beta \lambda } g^{\sigma \mu } p_1\cdot p_2-4 g^{\alpha
   \lambda } g^{\beta \nu } g^{\sigma \mu } p_1\cdot p_2+3
   g^{\alpha \beta } g^{\nu \lambda } g^{\sigma \mu } p_1\cdot
   p_2+3 g^{\beta \alpha } g^{\nu \lambda } g^{\sigma \mu }
   p_1\cdot p_2-4 g^{\alpha \mu } g^{\beta \lambda } g^{\sigma \nu
   } p_1\cdot p_2
  \right.  \nonumber\\
   && \left.
   +3 g^{\alpha \beta } g^{\mu \lambda } g^{\sigma
   \nu } p_1\cdot p_2+3 g^{\beta \alpha } g^{\mu \lambda }
   g^{\sigma \nu } p_1\cdot p_2\right) ;
\label{Thh}
\end{eqnarray}

\end{itemize}

\medskip

Two-loop integral appearing in results of various two-loop diagrams:
\begin{eqnarray}
  && \int\frac{d^n k_1 d^n k_2}{(2 \pi)^{2n}} \frac{1}{(k_1^2-M^2+i \epsilon)^\alpha
    (k_2^2-M^2+i \epsilon)^\beta ((k_1-k_2)^2+i \epsilon)^\gamma} = \nonumber \\
&& \qquad \qquad \ \ \ \frac{ i^{2-2 \alpha -2 \beta -2
   \gamma} M^{2 (n -\alpha -\beta
   -\gamma)} \Gamma \left(\frac{n}{2}-\gamma \right) \Gamma
   \left(-\frac{n}{2}+\alpha +\gamma \right) \Gamma
   \left(-\frac{n}{2}+\beta +\gamma \right)  \Gamma (-n+\alpha +\beta +\gamma )}{(4\pi)^n\Gamma (\alpha
   ) \Gamma (\beta ) \Gamma \left(\frac{n}{2}\right) \Gamma
   (-n+\alpha +\beta +2 \gamma )} \,.
\label{int2loop}
\end{eqnarray}

\end{document}